\documentclass[twocolumn,showpacs]{revtex4}

\usepackage{amsmath}
\usepackage{graphicx}
\usepackage{dcolumn}
\usepackage{bm}
\usepackage{amsfonts}
\usepackage{amssymb}
\begin{document}

\preprint{ITP/UU-XXX}

\title{Population and mass imbalance in atomic Fermi gases.}

\author{J. E. Baarsma}
\email{J.E.Baarsma@uu.nl}
\author{K. B. Gubbels}
\author{H. T. C. Stoof}

\affiliation{
Institute for Theoretical Physics, Utrecht University,\\
Leuvenlaan 4, 3584 CE Utrecht, The Netherlands}


\begin{abstract}
We develop an accurate theory of resonantly interacting Fermi
mixtures with both spin and mass imbalance. We consider Fermi
mixtures with arbitrary mass imbalances, but focus in particular
on the experimentally available ${}^{6}$Li-${}^{40}$K mixture. We
determine the phase diagram of the mixture for different
interactions strengths that lie on the BCS side of the Feshbach
resonance. We also determine the universal phase diagram at
unitarity. We find for the mixtures with a sufficiently large mass
imbalance, that includes the ${}^{6}$Li-${}^{40}$K mixture, a
Lifshitz point in the universal phase diagram that signals an
instability towards a supersolid phase.
\end{abstract}


\maketitle

\section{Introduction}
Ultracold quantum gases of fermionic atoms are at the center of
attention of both experimental and theoretical physicists. Because
of the amazing experimental control in these gases they offer the
possibility to experimentally explore various pairing phenomena.
As a result, many fundamental discoveries have already been made,
for example the realization of the crossover from a Bardeen-Cooper-Schrieffer (BCS) superfluid
of loosely bound Cooper pairs to a Bose-Einstein condensate (BEC)
of tightly bound molecules, also called the BEC-BCS crossover

For high-$T_{c}$ superconductors, the size of a Cooper pair is
comparable to the average distance between the electrons. This is
analogous to the intermediate regime in the BEC-BCS crossover. In
the BCS regime, the distance between two particles making up a
Cooper pair is much larger than the average distance between
particles, whereas in the BEC regime the distance between
particles within a pair is much smaller than the average
interparticle distance. If the size of a pair can be manipulated,
it is possible to go smoothly from one regime to the other. This
manipulation can be achieved in an atomic Fermi gas, where the
interaction strength between the atoms in the two different spin
states can be controlled with a Feshbach resonance. The smooth
BEC-BCS crossover was eventually realized in a trapped gas of
${}^{40}$K \cite{Regal} and ${}^{6}$Li atoms
\cite{Zwierlein,Kinast,Bartenstein,Bourdel,Partridge}.

For a Fermi mixture pairing is always possible for an equal amount
of particles in each spin state. However, pairing is absent for
the noninteracting system with all particles in one spin state.
Therefore, as a function of population imbalance there must exist
a phase transition at low temperatures. Experimentally, this
transition was studied for the strongly interacting and
mass-balanced case \cite{Ketterle,Hulet}, and the phase diagram
was found to be governed by a tricritical point that resulted in
the observation of phase separation \cite{Hulet,Shin}. Sarma superfluidity is also
likely to be present in this system \cite{Gubbels,Diederix}, but
has not been unambiguously identified yet.

Besides an imbalance in the particle densities, there can also be
an imbalance between the masses of the particles. Consequences of
these imbalances are interesting to study, also because imbalanced
Fermi mixtures exist in certain condensed-matter systems in a
magnetic field, in nuclear matter, and even in the quark-gluon
plasma that is supposed to be present in the core of heavy neutron
stars \cite{Bailin,Casalbuoni}.

While studying mass-imbalanced Fermi mixtures, we first focused on
the mixture with a mass ratio of 6.7, corresponding to a
${}^6$Li-${}^{40}$K mixture. This resulted in a Letter \cite{prl}
where we presented the phase diagram of this mixture as a function
of polarization and temperature in the strongly interacting limit,
or the so-called unitarity limit, where the scattering length of
the interparticle interaction diverges. In the unitarity limit,
the size of the Cooper pairs is comparable to the average
interparticle distance and the pairing is a many-body effect. As a
result, the mass imbalance has profound effects on the pairing,
because the imbalance strongly affects the two Fermi spheres that
are present in the system. We showed that the phase diagram of the
${}^6$Li-${}^{40}$K mixture contains a Lifshitz point for a
majority of heavy fermions \cite{prl}. Typically, Lifshitz points
are found at weak interactions where the critical temperatures are
very low. However, we found that in the strongly interacting limit
the phase diagram of the ${}^6$Li-${}^{40}$K mixture already
contains a Lifshitz point at accessible temperatures, see
Fig.~\ref{r1 met fluctuaties} below. At a Lifshitz point the phase transition
undergoes a dramatic change of character. Rather than preferring a
homogeneous order parameter, the system now forms an inhomogeneous
superfluid.

The possibility of an inhomogeneous superfluid was early
investigated by Larkin and Ovchinnikov (LO), who considered a
superfluid with a single standing-wave order parameter
\cite{Larkin}, which is energetically more favorable than the
plane-wave case studied by Fulde and Ferrell (FF)\cite{Fulde}.
Since the LO phase results in periodic modulations of the particle
densities, it is a supersolid \cite{Bulgac}. The FF and LO phases
have intrigued the condensed-matter community for many decades,
but only very recently strong evidence for the FFLO phase has been
obtained in a one-dimensional imbalanced Fermi mixture of two spin
states \cite{Liao}. Theoretically, it is challenging to describe
the phase diagram below a Lifshitz point. The most stable states
are likely to be complicated superpositions of standing waves,
where different ansatzes lead to different stability regions
\cite{Mora,Yip}.

Initially, our focus was on the ${}^6$Li-${}^{40}$K
mixture, because this is experimentally a very promising mixture,
where several accessible Feshbach resonances are identified
\cite{Walraven} and both species have also been simultaneously
cooled into the degenerate regime \cite{Dieckmann}. In contrast to
our previous work, we now consider Fermi mixtures with an
arbitrary mass ratio. Moreover, apart from the unitarity limit, we
also cover the BCS regime. We do not calculate phase diagrams on
the BEC side of the Feshbach resonance, because in that regime we
should include thermal molecules in our calculations, which
requires a different theory. Due to the population and the mass
imbalance, the two Fermi spheres in the system are typically
mismatched, which can induce phase separation with the location of
the tricritical point depending on the mass ratio and the
interaction strength \cite{Parish}. In this paper, we also
consider Lifshitz instabilities and also find a multicritical
point for the mass-balanced case at a very weak interaction, as
shown in Fig.~\ref{tc3}. At this interaction strength the Lifshitz
point and the tricritical point occur for the same polarization
and temperature. Furthermore, in the imbalanced Fermi mixture the
quasiparticle dispersions in the superfluid phase also give rise
to gapless Sarma superfluidity. For the mass-balanced Fermi
mixture and for the ${}^6$Li-${}^{40}$K mixture we therefore also
calculated the regions in the phase diagrams where the superfluid
is gapless.

In our Letter \cite{prl}, we showed that although mean-field theory vastly overestimates critical temperatures in the unitarity limit, it is very useful for a qualitative description of the physics. From the mass-balanced case, we know that the
critical temperatures found using mean-field techniques are lowered mainly by two effects, namely
the fermionic selfenergies and the screening of the interaction
due to particle-hole fluctuations \cite{Gorkov,Koos}. After
discussing mean-field theory for the imbalanced Fermi gases, we
take both these effects into account in the unitarity limit for
the mass-balanced case and for the experimentally interesting case
of the ${}^6$Li-${}^{40}$K mixture. This leads to results that
compare well with Monte Carlo calculations \cite{Burovski} and for
equal masses also with experiment \cite{Shin}. Our procedure gives
a reduction of the mean-field critical temperatures by a factor of
3. This makes it experimentally more difficult, but not
impossible, to reach also for the mass-imbalanced case the
superfluid regime. Very importantly, the Lifshitz point remains
present in the phase diagram after taking fluctuations into
account. This hopefully brings the observation of inhomogeneous
superfluidity within experimental reach.

This paper is organized as follows. We start with discussing the
interactions in Fermi mixtures in Sec.~II. In Sec.~III, we give a
brief discussion of the Landau theory that we use to describe
phase transitions. In particular, we introduce the Landau
thermodynamic potential and the order parameter. Next, in Sec.~IV,
we discuss the mean-field theory that we use to calculate phase
diagrams, which are presented in Sec.~IVC. We discuss the phase
diagrams for three different mass ratios at different interaction
strengths, to explore the various topologies of the phase diagrams
that can arise. After that, we discuss in more detail the effects
of the mass imbalance. We then also explore the effect of the
interaction strength on the position of the tricritical points and
Lifshitz points in the phase diagram. Subsequently, we discuss the
presence of the superfluid Sarma phase for the mass-balanced case
and the ${}^6$Li-${}^{40}$K mixture, both at unitarity. All these
calculations use in first instance mean-field theory. Then, in
Sec.~V, we include fluctuation effects to obtain more
quantitative results for the mass-balanced case and the ${}^6$Li-${}^{40}$K mixture. We focus here on the unitarity limit, although fluctuation effects could also be easily incorporated in
the BCS limit of the Feshbach resonance. Finally, different
appendices are added where more calculations can be found. In
Appendix A it is explained how the thermodynamic potential for the
mass-imbalanced Fermi gas is obtained, while Appendix B contains
the calculations for the amplitudes of relevant Feynman diagrams.

\section{Interactions and Feshbach resonances}

In this paper we study phase transitions in an imbalanced Fermi
gas at different interaction strengths. This is in particular
relevant if the interaction strength is experimentally under
control. In atomic Fermi mixtures the interspecies interaction can
be controlled using a Feshbach resonance \cite{Duine}. The effect of the
microscopic interaction potential can be studied via the two-body
transition operator $\hat{T}^{\text{2B}}$. The matrix elements of
this transition operator are directly related to the scattering
amplitudes. It is defined by
\begin{align}
\hat{V}|\mbox{\boldmath$\psi$}^{(+)}_{\mathbf{k}}\rangle
  \equiv\hat{T}^{\text{2B}}|\mathbf{k}\rangle,
\end{align}
where $|\mbox{\boldmath$\psi$}^{(+)}_{\mathbf{k}}\rangle$ are the
scattering states and $\hat{V}$ is the microscopic interaction
potential. To find an expression for $\hat{T}^{\text{2B}}$ we
start with the Lippmann-Schwinger equation
\begin{align}
\hat{T}^{\text{2B}}=\hat{V}+\hat{V}\frac{1}{z-\hat{H}_{0}}\hat{T}^{\text{2B}},
\label{lippmannschwinger}
\end{align}
where $z=E+\text{i}0$ and the notation $\text{i}0$ implies the
limit $\text{i}\varepsilon$ with $\varepsilon\downarrow 0$. In
this paper we study an imbalanced Fermi gas with a point
interaction
\begin{align}
V(\mathbf{x}-\mathbf{x'})\simeq
V_{0}\delta(\mathbf{x}-\mathbf{x'}),
\end{align}
where $V_{0}$ is negative, since we are interested in an
attractive interaction. If we now consider the Lippmann-Schwinger
equation at zero energy $z=0$, multiply both sides with
$\langle\mathbf{k'}|$ from the left and with
$|\mathbf{k''}\rangle$ from the right, and we insert a
completeness relation in the second term on the right-hand side,
we obtain
\begin{align}
\frac{1}{T^{\text{2B}}(0)}=\frac{1}{V_{0}}
+\int\frac{\text{d}\mathbf{k}}{(2\pi)^{3}}
\frac{1}{2\varepsilon(\mathbf{k})}, \label{t2b bij puntinteractie}
\end{align}
where half the reduced kinetic energy,
$\varepsilon(\mathbf{k})=\hbar^{2}\mathbf{k}^{2}/2m$, is the
kinetic energy associated with a mass $m$ that is equal to twice
the reduced mass, namely
\begin{equation}
\frac{2}{m}=\frac{1}{m_{+}}+\frac{1}{m_{-}}. \label{twice}
\end{equation}
Here, $m_{+}$ and $m_{-}$ are the masses of a light and a heavy particle, respectively.

\begin{figure}
\includegraphics[width=.45\textwidth]{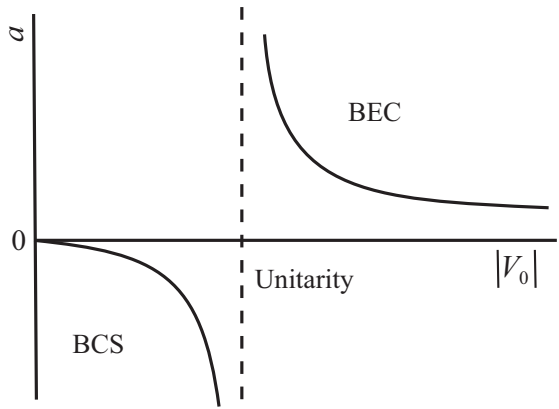}
\caption{The $s$-wave scattering length $a$ as a function of the
microscopic interaction strength $V_{0}$. Here $V_{0}$ is
negative, since we are considering an attractive interaction.}
\label{fesh}
\end{figure}

The two-body transition matrix is related to the $s$-wave
scattering length $a$ by
\begin{align}
\frac{1}{T^{\text{2B}}(0)}=\frac{m}{4\pi\hbar^{2}a}. \label{a}
\end{align}
Since we are interested in the behavior of the Fermi gas at
ultralow temperatures, we can use a cut-off momentum
$\hbar\Lambda$ to evaluate the integral in Eq.~(\ref{t2b bij
puntinteractie}) and we obtain, using Eq.~(\ref{a}), a relation
between the scattering length $a$ and the microscopic interaction
potential $V_{0}$, namely
\begin{align}
a=\frac{m\pi V_{0}}{2 m\Lambda V_{0}+4\pi^2\hbar^{2}}.
\end{align}
This relation is shown in Fig.~\ref{fesh}. It can be seen that for
small values of $V_{0}$ the scattering length is negative, which
means that the Fermi mixture is in the BCS regime. Then, for
$V_{0}=-(2\pi\hbar)^{2}/m\Lambda$ the scattering length diverges,
which is called the unitarity limit. For large values of $V_{0}$
the scattering length is positive and the Fermi mixture is in the
BEC regime.

For $a>0$ the microscopic interaction has a bound state with an
energy $E=-\hbar^{2}/ma^{2}$. This is the single-channel picture
of a Feshbach resonance. In this paper we focus on the unitarity
limit and on the BCS side of the resonance. Thus, we look at the
case where $a$ diverges and at negative scattering lengths $a$.
For those scattering lengths we can use the single-channel
picture, as long as the Feshbach resonance is sufficiently broad.
Namely, in the limit of a broad resonance the amplitude to be in
the bare molecular state of the Feshbach resonance turns out to be
very small \cite{Romans,Partridge}.

\section{Landau Theory of Phase Transitions}\label{landautheory}

In order to study the critical behavior of a system, in our case
an imbalanced Fermi gas, we consider the Landau thermodynamic
potential density $\omega_{\text{L}}(\Delta(\mathbf{x}))$, with
$\Delta(\mathbf{x})$ the superfluid order parameter. Near the
phase transition, where the BCS order parameter
$\Delta(\mathbf{x})$ is small, the Landau thermodynamic potential
density can be expanded as \cite{Negele,uqf}
\begin{equation}
\omega_{\text{L}}(\Delta;\mu_{\sigma},T)=\gamma|\nabla\Delta|^{2}+
\alpha|\Delta|^{2}+\frac{\beta}{2}|\Delta|^{4}+\ldots,
\end{equation}
where the dots denote the higher orders in $|\Delta|^{2}$ and in
gradients $|\nabla\Delta|^{2}$. The Landau coefficients in the
thermodynamic potential all depend on the temperature and on the
chemical potentials of the two fermion species. If in the
thermodynamic potential all coefficients are positive, the minimum
of the thermodynamic potential is located at
$\langle\Delta(\mathbf{x})\rangle$ equal to zero and the system
will be in the normal state. Whereas a phase transition to a
superfluid state has occurred when the position of the global
minimum is located at a nonzero order parameter
$\langle\Delta(\mathbf{x})\rangle$, which describes a condensate
of bosonic pairs. In the case that $\gamma$ is positive, it costs
energy to have a spatially varying superfluid. It is then
energetically favorable for the system to be homogeneous and
therefore we can restrict ourselves to a pairing field $\Delta$
independent of position.

\begin{figure}
\includegraphics[width=\columnwidth]{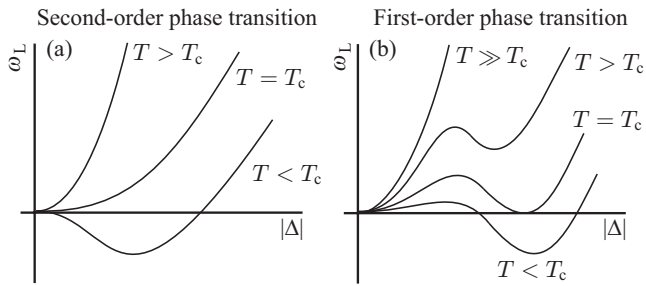}
\caption{The thermodynamic potential density
$\omega_{\text{L}}(|\Delta|)$ as a function of the order parameter
$\Delta$. Panel (a) shows the behavior of
$\omega_{\text{L}}(|\Delta|)$ for different temperatures when a
second-order phase transition occurs and panel (b) when a
first-order phase transition occurs.} \label{fases}
\end{figure}

We consider first the case where $\gamma$ is positive and the
system is homogeneous. For high temperatures all coefficients in
the thermodynamic potential will be positive and the system will
be in the normal state. But for low temperatures it can occur that
certain coefficients change sign. Suppose that in the Landau
thermodynamic potential $\alpha$ is negative and all other
coefficients are positive. The minimum of the thermodynamic
potential will then be attained at some nonzero
$\langle\Delta\rangle$ and the Fermi gas will be in the superfluid
state. Thus, as $\alpha$ changes sign a phase transition takes
place. The temperature at which the transition from the normal
state to the superfluid state occurs, for given chemical
potentials, can be determined by equating the quadratic
coefficient to zero, $\alpha(T_{\text{c}})=0$, where
$T_{\text{c}}$ is called the critical temperature. The phase
transition just described is called a second-order phase
transition and it is characterized by the fact that the minimum of
the thermodynamic potential shifts away from zero continuously,
see Fig.~\ref{fases}a.

It is also possible to have a first-order phase transition. To
explain a first-order phase transition we consider the situation
where in the thermodynamic potential density all coefficients but
the fourth-order coefficient $\beta$ are positive. The
thermodynamic potential will then typically have two minima. One
of these is located at $\Delta$ equal to zero and the other one
will be located at a nonzero value of the order parameter. For
higher temperatures the minimum located at zero is a global
minimum and the Fermi gas is in the normal state. If the
temperature is lowered there will be a point where the two minima
are equal and for even lower temperatures the minimum located at a
nonzero order parameter $\langle\Delta\rangle$ is the global
minimum. The system is then in the superfluid state. The
first-order phase transition takes place when these two minima are
equal, i.e., when
$\omega_{\text{L}}(0)=\omega_{\text{L}}(\langle\Delta\rangle)$. In
contrast to a second-order phase transition, the location of the
global minimum of the thermodynamic potential now changes
discontinuously from being zero to a nonzero
$\langle\Delta\rangle$, see Fig.~\ref{fases}b.

Next, we consider the case where $\gamma$ is negative. The system
can then gain energy when the order parameter varies in space.
Thus, instead of being constant, the order parameter will now
depend on position. Fulde and Ferrell studied the plane-wave
solution \cite{Fulde}
\begin{equation}
\Delta(\mathbf{x})=\Delta_{0}e^{\text{i}\mathbf{k\cdot x}},
\end{equation}
while Larkin and Ovchinnikov considered a superfluid with the
single standing-wave order parameter \cite{Larkin}
\begin{equation}
\Delta(\mathbf{x})=\Delta_{0}\cos(\mathbf{k\cdot x}),
\end{equation}
which turns out to be energetically more favorable than the
plane-wave case. Superpositions of more than two plane waves are
also possible.

In the LO phase the wavefunction of the bosonic pairs is periodic and therefore there will also exist a periodicity in the atomic density. This periodic structure shows itself in the diagonal elements of the one-particle density matrix $n(\mathbf{x},\mathbf{x'})$ and is therefore called diagonal long-range order. This diagonal long-range order is what characterizes a solid. In a fermionic superfluid a fraction of the Cooper pairs is in the lowest energy eigenstate. There thus exists a long-range order between the positions of the pairs. It is also said that the two-particle density matrix $g(\mathbf{x},\mathbf{x'})$ has off-diagonal long-range order, which implies that $g(\mathbf{x},\mathbf{x'})$ does not vanish in the limit $|\mathbf{x}-\mathbf{x'}|\rightarrow\infty$ and characterizes a superfluid. In the case where $\gamma$ is negative a transition occurs from the normal state to a superfluid where the atomic density in the superfluid has a periodic structure. Then there exists both diagonal long-range order as well as off-diagonal long-range order and the state is both solid and superfluid. This is called a supersolid phase \cite{Bulgac,uqf}.

Further on, we show that in a mass-imbalanced Fermi gas not only a
second-order phase transition can occur, but also a first-order
transition and even a transition to an inhomogeneous superfluid,
depending on the values of the two chemical potentials. Therefore,
there must be points in the phase diagram where the character of
the phase transition changes. If the phase transition changes from
being second order to first order there will be a tricritical
point \cite{Chaikin}. This point can thus be found by setting
\begin{equation}
\alpha(T_{\text{c}3})=\beta(T_{\text{c}3})=0, \label{tricrit}
\end{equation}
where $T_{\text{c}3}$ is the tricritical temperature. For
temperatures higher than the tricritical temperature the phase
transition from the normal to the superfluid state is of second
order, while for $T<T_{\text{c}3}$ there is a first-order phase
transition and phase separation occurs.

The phase transition could also change from being a transition
from the normal state to a homogeneous superfluid to a transition
from the normal state to a supersolid. The point in the phase
diagram where this occurs is called a Lifshitz point \cite{Chaikin}. This point
can be computed by demanding
\begin{equation}
\alpha(T_{\text{L}})=\gamma(T_{\text{L}})=0,\label{lifshitz}
\end{equation}
where $T_{\text{L}}$ is called the Lifshitz temperature. For
temperatures lower than the Lifshitz temperature the transition
will be from a normal state to a superfluid state where the
bosonic pairs have a nonzero momentum. Superfluidity at nonzero
momentum can be established in many ways and due to this variety
of possibilities it is difficult to predict which kind of
superfluidity will be present below the Lifshitz point. However,
they all have to emerge from the Lifshitz point and therefore it
is important to know the position of the Lifshitz point.

Apart from the above two possibilities, we can think of other
scenarios for the change in character of the phase transitions.
But in the phase diagrams we calculated for the imbalanced Fermi
gas we only found tricritical points and Lifshitz points.
Therefore, these are the only two possibilities we discuss in the
following.

\section{Mean-Field Theory}

In this section we present the mean-field thermodynamic potential,
which is an approximation to the exact Landau thermodynamic
potential for the Fermi gas with population and mass imbalance. If
we have an expression for the Landau thermodynamic potential, we
are able to determine the phase diagram.

Although mean-field theory does not contain all interactions present in a Fermi mixture, it turns out that mean-field theory already incorporates all the relevant physics determining the topology of the phase diagrams. Adding fluctuation effects only changes the phase diagrams quantitatively \cite{prl,Koos,Gorkov,Iskin}. Because of the rather straightforward and transparent calculations in mean-field theory, we first discuss mean-field theory in some detail. Later on we take fluctuation effects into account in order to obtain more quantitative results which we can compare with Monte Carlo calculations and for the mass-balanced case with experiment.

\subsection{Thermodynamic Potential}

We consider a two-component Fermi mixture, i.e., a mixture
containing either a single fermionic species, for which two
different hyperfine states are present, or consisting of two
different fermionic species with access to a single hyperfine
state. A balanced Fermi gas consists of a single species with an
equal population of both spin states. In an imbalanced Fermi gas
we allow the populations and masses to be different. To take into
account a population imbalance we use different chemical
potentials for the two (pseudo)spin states, while a mass imbalance
implies that the particles in the two hyperfine states
have different masses. The chemical potential and the mass of the
fermions in state $|\sigma\rangle$ will be denoted by
$\mu_{\sigma}$ and $m_{\sigma}$ respectively. From now on heavy particles
are always denoted by a minus sign and light particles by a plus
sign, thus $\sigma=\pm$.

The mean-field thermodynamic potential for the imbalanced Fermi
gas is given by
\begin{eqnarray}
\nonumber\omega_{\text{L}}(|\Delta|)=-\frac{|\Delta|^{2}}{T^{\text{2B}}(0)}
+\int\frac{\text{d}\mathbf{k}}{(2\pi)^{3}}\Big\{
\varepsilon(\mathbf{k})-\mu-\hbar\omega(\mathbf{k})\\
+\frac{|\Delta|^{2}}{2\varepsilon(\mathbf{k})}
-\frac{1}{\beta}\sum_{\sigma}\log\left(1+e^{-\beta\hbar\omega_{\sigma}(\mathbf{k})}\right)\Big\},
\label{thermpot}
\end{eqnarray}
where the two-body transition matrix $T^{\text{2B}}(0)$ is given
by Eq.~(\ref{a}). This thermodynamic potential is a direct
generalization of the thermodynamic potential for a balanced Fermi
gas \cite{uqf,Fetter}. In the above expression $\beta=1/k_{\text{B}}T$ is the
inverse thermal energy, $\mu=(\mu_{+}+\mu_{-})/2$ is the average
chemical potential and half the reduced kinetic energy is
$\varepsilon(\mathbf{k})=(\varepsilon_{+}(\mathbf{k})+\varepsilon_{-}(\mathbf{k}))/2$
with
\begin{eqnarray}
\varepsilon_{\sigma}(\mathbf{k})=\frac{\hbar^{2}\mathbf{k}^{2}}{2
m_{\sigma}}. \label{energies}
\end{eqnarray}
The first terms in the above thermodynamic potential represent the
BCS ground state of the mixture, where $\hbar\omega(\mathbf{k})$
is the average dispersion of the quasiparticles \begin{equation}
\hbar\omega(\mathbf{k})=\sqrt{(\varepsilon(\mathbf{k})-\mu)^{2}+|\Delta|^{2}},
\label{dispersie}
\end{equation}
with $|\Delta|$ the so called BCS gap parameter. The complex
pairing field $\Delta$ is on average related to the expectation
value of the pair annihilation operator through
\begin{equation}
\left<\Delta(\mathbf{x})\right>=V_{0}\left<\hat{\psi}_{-}(\mathbf{x})\hat{\psi}_{+}(\mathbf{x})\right>,
\label{bosonic fields}
\end{equation}
with $\hat{\psi}_{\sigma}(\mathbf{x})$ the fermionic annihilation
operators.

\begin{figure}
\includegraphics[width=\columnwidth]{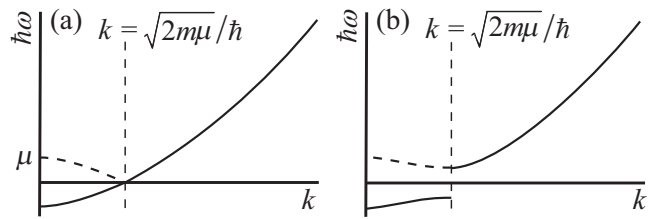}
\caption{The single-particle dispersions
$\hbar\omega_{\sigma}(\mathbf{k})$ for the balanced Fermi gas.
Panel (a) shows the dispersions for zero $\Delta$ where the system
is in the normal state. Panel (b) depicts the case where the system
is in the superfluid state, thus the dispersions for nonzero
$\Delta$. In both panels the dashed lines are the dispersions of
the hole-like excitations, while the full lines give the
particle-like dispersions. The latter show more clearly the
opening of a gap $2\Delta$ at the Fermi level due to the formation
of a condensate of Cooper pairs.} \label{dispersionsbalanced}
\end{figure}

\begin{figure}
\includegraphics[width=\columnwidth]{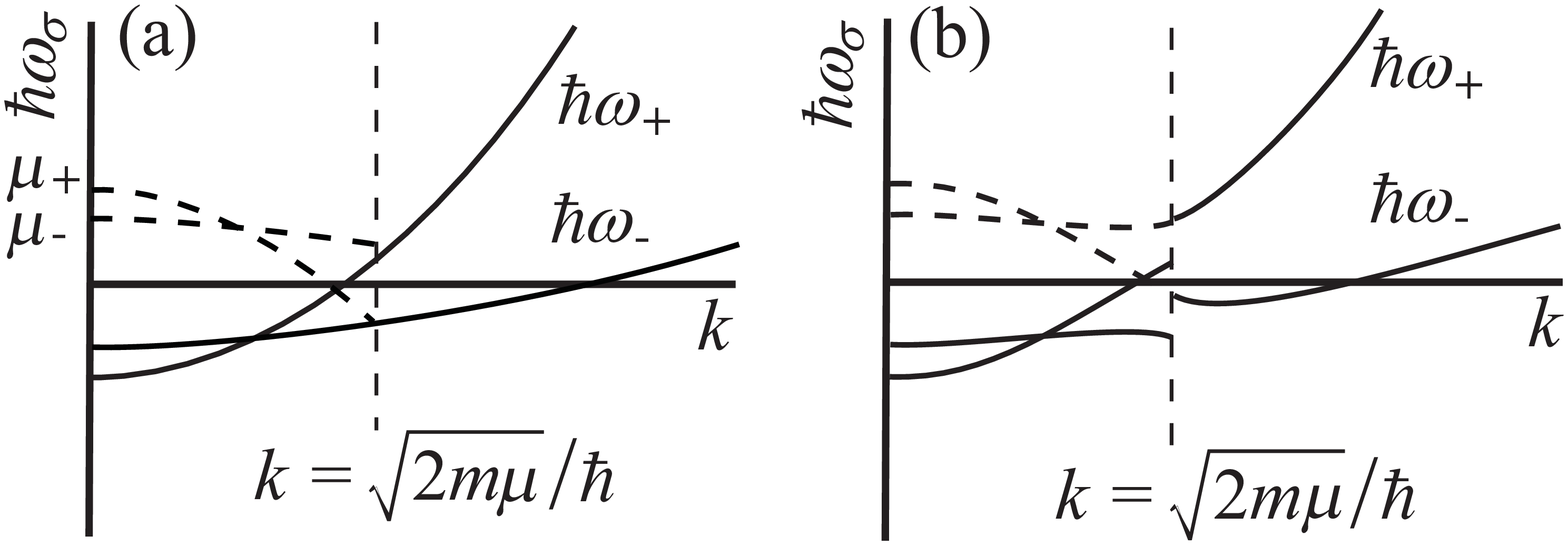}
\caption{The dispersions $\hbar\omega_{\sigma}(\mathbf{k})$ for
the Fermi gas with both population and mass imbalance. Panel (a)
shows again the dispersions for zero $\Delta$ where the system is
in the normal state. And panel (b) again depicts the case where the
system is in the superfluid state, thus the dispersions for
nonzero $\Delta$. The meaning of the dashed and full lines are the
same as for Fig. \ref{dispersionsbalanced}.}
\label{dispersionsimbalanced}
\end{figure}

The second part of the thermodynamic potential, namely the part
containing the logarithms, corresponds to the contribution of an
ideal gas of quasiparticles. Here,
$\hbar\omega_{\sigma}(\mathbf{k})$ is the dispersion relation of
the quasiparticles in state $|\sigma\rangle$, given by
\begin{equation}
\hbar\omega_{\sigma}(\mathbf{k})=\hbar\omega(\mathbf{k})-\sigma[2h-
\varepsilon_{+}(\mathbf{k})+\varepsilon_{-}(\mathbf{k})]/2,
\label{dispersieplus}
\end{equation}
with $h=(\mu_{+}-\mu_{-})/2$ the difference in chemical
potentials. For the unpolarized Fermi gas with equal masses the
dispersions $\hbar\omega_{\sigma}(\mathbf{k})$ reduce to the
average dispersion in Eq.~(\ref{dispersie}). This dispersion is
plotted in Fig.~\ref{dispersionsbalanced} for both the normal
state (Fig.~\ref{dispersionsbalanced}a) and the superfluid state
(Fig.~\ref{dispersionsbalanced}b). In this case the superfluid is
gapped and balanced. For $k=|\mathbf{k}|<\sqrt{2m\mu}/\hbar$ the
quasiparticle dispersion describes hole-like excitations. If we
mirror this hole-like part of the quasiparticle dispersion, we
obtain the negative dispersion of the particle-like excitations.
For $k>\sqrt{2m\mu}/\hbar$ the quasiparticle dispersion already describes
the particle-like excitations.

In the case of a polarized superfluid with equal masses the
dispersions $\hbar\omega_{\sigma}(\mathbf{k})$ in
Fig.~\ref{dispersionsbalanced} are shifted by the difference in
chemical potentials $2h$. For $h>|\Delta|$ the dispersion of the
majority species becomes negative. When this occurs, the occupation
of the single-particle states associated with the negative part of
the quasiparticle excitation branch actually lowers the ground-state energy. Since this leads to additional majority
quasiparticles and, therefore, additional majority particles and minority holes, in
the ground state, the ground state becomes a polarized superfluid.
The resulting gapless and polarized superfluid is called the Sarma
phase \cite{Sarma}. For the Fermi gas with both mass and population imbalance
the dispersions are depicted in Fig.~\ref{dispersionsimbalanced},
where Fig.~\ref{dispersionsimbalanced}a shows the dispersions for
zero $\Delta$ and Fig.~\ref{dispersionsimbalanced}b for nonzero
$\Delta$. The shape of the dispersions is changed due to the
difference in mass. It can be seen that in this case also one of
the dispersions, namely $\hbar\omega_{-}(\mathbf{k})$, is negative
so that it would correspond to a gapless Sarma superfluid.

\subsection{Landau Coefficients}

For the imbalanced Fermi mixture, we want to study the phase
transition from the normal state to the superfluid state. For that
transition we want to obtain a phase diagram with the critical
temperature as a function of the polarization,
$P=(n_{+}-n_{-})/(n_{+}+n_{-})$ with $n_{\sigma}$ the density of
particles in state $|\sigma\rangle$.

\begin{figure}
\includegraphics[width=.4\textwidth]{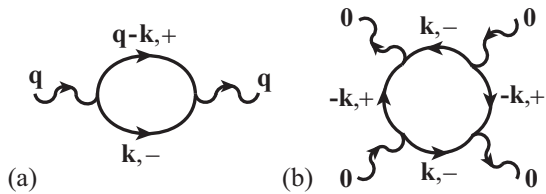}
\caption{(a) The ladder diagram with external momentum
$\mathbf{q}$. The wiggly lines denote the Cooper pairs, which can
break up into two fermions of different spin, denoted by the
normal lines. Here, $\mathbf{q}$ is the wavevector of the Cooper
pairs, while $\mathbf{k}$ and $\mathbf{q}-\mathbf{k}$ are the
wavevectors of the fermions. When $\mathbf{q}$ is equal to zero,
the amplitude of this diagram corresponds to the quadratic Landau
coefficient $\alpha$ in the thermodynamic potential. (b) The
Feynman diagram corresponding to the fourth-order Landau
coefficient $\beta$.} \label{ladderq}
\end{figure}

The particle densities $n_{\sigma}$ can be determined from the
thermodynamic potential using \cite{uqf}
\begin{equation}
n_{\sigma}=\left.-\frac{\partial\omega_{\text{L}}(\Delta)}{\partial\,\mu_{\sigma}}\right|_{\Delta=\langle\Delta\rangle}.
\label{particle densities}
\end{equation}
Moreover, in Sec.~\ref{landautheory} the critical conditions
for the transition were explained. In mean-field theory, the
quadratic Landau coefficient $\alpha$ is explicitly given by
\begin{align}
\nonumber&\alpha=\left.\frac{\partial\omega_{\text{L}}}{\partial|\Delta|^{2}}\right|_{|\Delta|^{2}=0}=-\frac{1}{T^{\text{2B}}(0)}\\
&+\int\frac{\text{d}\mathbf{k}}{(2\pi)^{3}}\left(\frac{1}{2\varepsilon(\mathbf{k})}+\frac{N_{+}(\mathbf{k})+N_{-}(\mathbf{k})-1}{2(\varepsilon(\mathbf{k})-\mu)}\right),
\label{alpha}
\end{align}
where
$N_{\sigma}(\mathbf{k})=1/(\exp[\beta(\varepsilon_{\sigma}(\mathbf{k})-\mu_{\sigma})]+1)$
are the Fermi distribution functions. To determine the temperature
of the tricritical point, we also need to know the fourth-order
Landau coefficient $\beta$. It is given by
\begin{eqnarray}
\nonumber\beta=\left.\frac{\partial^{2}\omega_{\text{L}}}{(\partial|\Delta|^{2})^{2}}\right|_{|\Delta|^{2}=0}=\int\frac{\text{d}^{3}\mathbf{k}}{(2\pi)^{3}}\frac{1}{4(\varepsilon(\mathbf{k})-\mu)^{2}}\\
\nonumber\noindent\times\Bigg[\beta N_{+}(\mathbf{k})(N_{+}(\mathbf{k})-1)+\beta N_{-}(\mathbf{k})(N_{-}(\mathbf{k})-1)\\
\noindent\hspace{4mm}+\frac{1}{\varepsilon(\mathbf{k})-\mu}(1-N_{+}(\mathbf{k})-N_{-}(\mathbf{k}))\Bigg].
\label{beta}
\end{eqnarray}
Determining $\gamma$ from the mean-field thermodynamic potential
is not possible, since we have assumed the bosonic pairing field
$\Delta(\mathbf{x})$ to be independent of position. Nevertheless,
there is a rather simple way to determine this coefficient, using
Feynman diagrams \cite{Kleinert}. The other coefficients could also have been
determined using a diagrammatic language. Namely, the quadratic
coefficient $\alpha$ corresponds to the so called ladder diagram
where the incoming and outgoing bosonic fields $\Delta$ have zero
momentum, see Fig.~\ref{ladderq}a. Physically, $\alpha$ can be
interpreted as being proportional to the chemical potential of the
Cooper pairs. The fourth-order coefficient $\beta$ has a
diagrammatic representation with four external bosonic fields with
zero momentum, see Fig.~\ref{ladderq}b.

\begin{figure*}[t]
\includegraphics[width=0.9\textwidth]{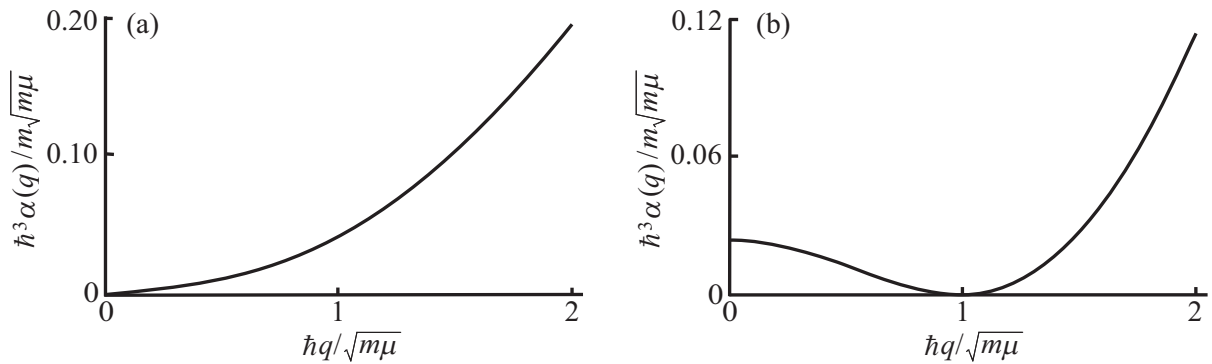}
\caption{Depending on the temperature and the chemical potentials,
the minimum of $\alpha(\mathbf{q})$ is either at zero or at
nonzero external momentum $\hbar q$. In both pictures the mass
ratio is $r=6.7$. In panel (a), the minimum of
$\alpha(\mathbf{q})$ is located at zero external momentum. The
corresponding polarization is $P=-0.61$ and the temperature is
$T/T_{\text{F}}=0.21$ with $T_{\text{F}}$ the Fermi temperature.
In panel (b), the minimum of $\alpha(\mathbf{q})$ is attained at a
nonzero momentum. Here the polarization is $P=-0.7$ and the
temperature is $T/T_{\text{F}}=0.13$.} \label{alphaq}
\end{figure*}

For the transition to the supersolid phase, we consider the ladder
diagram where the bosonic fields carry nonzero momentum
$\mathbf{q}$, since the supersolid phase consists of bosonic pairs
with nonzero momentum. The expression for this ladder diagram is
\begin{align}
\nonumber&\alpha(\mathbf{q})=-\frac{1}{T^{\text{2B}}(0)}+\int\frac{\text{d}^{3}\mathbf{k}}{(2\pi)^{3}}\Bigg\{\frac{1}{2\varepsilon(\mathbf{k})}+\\
&\frac{N_{+}(\mathbf{q-k})+N_{-}(\mathbf{k})-1}{\varepsilon_{+}(\mathbf{q-k})+\varepsilon_{-}(\mathbf{k})-2\mu}\Bigg\}.
\label{extern}
\end{align}

The actual shape of $\alpha(\mathbf{q})$ as a function of the
external momentum $\mathbf{q}$ depends on the values of the
different parameters in this expression, such as the chemical
potentials and the temperature. Depending on those quantities the
minimum of $\alpha(\mathbf{q})$ is attained either for zero or for
nonzero external momentum. As explained in Sec.~\ref{landautheory}, a second-order phase transition can occur,
when a quadratic coefficient of the Landau theory changes sign.
These coefficients are now given by $\alpha({\bf q})$, where
${\bf q}$ is the wavevector of the bosonic pairs. The sign change
occurs first for the minimum of $\alpha({\bf q})$, which therefore
determines whether or not the transition happens at nonzero ${\bf
q }$. The expression for the ladder diagram with nonzero external
momentum can be expanded in even powers of $\mathbf{q}$
\begin{equation}
\alpha(\mathbf{q})=a_{0}+a_{1}\mathbf{q}^{2}+a_{2}\mathbf{q}^{4}+\ldots,
\label{external}
\end{equation}
where the dots denote higher order powers in $\mathbf{q}^{2}$. If
all the coefficients $a_{i}$ are positive, $\alpha(\mathbf{q})$
has a minimum for external momentum zero, so that a transition to
the homogeneous superfluid phase occurs. But if $a_{1}$ is
negative, $\alpha(\mathbf{q})$ has a minimum for a nonzero
external momentum and therefore it first becomes zero at some
nonzero value of $\mathbf{q}$. The minimum of the thermodynamic
potential will then be located at a nonzero order parameter with a
nonzero momentum. In other words, it will be energetically
favorable for the bosonic pairs to have kinetic energy and the
phase transition that occurs is a transition from the normal state
to an inhomogeneous superfluid. Comparing this with the Landau
theory of Sec.~\ref{landautheory}, we see that in the expansion
of $\alpha(\mathbf{q})$, $a_{0}$ can be identified with the
quadratic coefficient $\alpha$ and $a_{1}$ can be identified with
$\gamma$. Thus, from the ladder diagram with external momentum an
expression for $\gamma$ can be found, namely
\begin{equation}\label{gamma}
\gamma=\left.\frac{\partial\alpha(\mathbf{q})}{\partial\mathbf{q}^{2}}\right|_{\mathbf{q}=0}.
\end{equation}
Physically, $\gamma$ can be interpreted as being proportional to
the inverse of the effective mass of the Cooper pairs.

\subsection{Results}

In this section the results using mean-field theory are presented.
First, we present phase diagrams for three Fermi gases with
different mass imbalances. Then, we study the effect of the
mass imbalance on the critical temperature for an unpolarized
Fermi gas. After this, we study the effect of the interaction
strength on the temperature corresponding to a tricritical point
or a Lifshitz point. Finally, we also consider the superfluid
Sarma phase.

\subsubsection{Phase Diagrams}

With the expressions for the Landau coefficients, the phase
diagram can be calculated for a fixed mass ratio $r$ and a fixed
interaction strength $1/k_{\text{F}}a$. We determine the phase
diagram as a function of temperature $T$ and polarization
$P=(n_{+}-n_{-})/(n_{+}+n_{-})$. The mass ratio $r$ is given by
$r=m_{-}/m_{+}$. The interaction strength is characterized by the
$s$-wave scattering length $a$ and the Fermi momentum
$k_{\text{F}}$, which is defined as
\begin{equation}
k_{\text{F}}=(3\pi^{2}n)^{1/3}, \label{fermimomentum}
\end{equation}
where $n=n_{+}+n_{-}$ is the total particle density. In order to
obtain a phase diagram independent of the total particle density
$n$, we scale the temperature with the reduced Fermi temperature
\begin{equation}
k_{\text{B}}T_{\text{F}}=\varepsilon_{\text{F}}=\frac{\hbar^{2}k_{\text{F}}^{2}}{2m},
\end{equation}
where $m$ is twice the reduced mass, introduced in
Eq.~(\ref{twice}) and $k_{\text{B}}$ is Boltzmann's constant.

\begin{figure*}
\includegraphics[width=\textwidth]{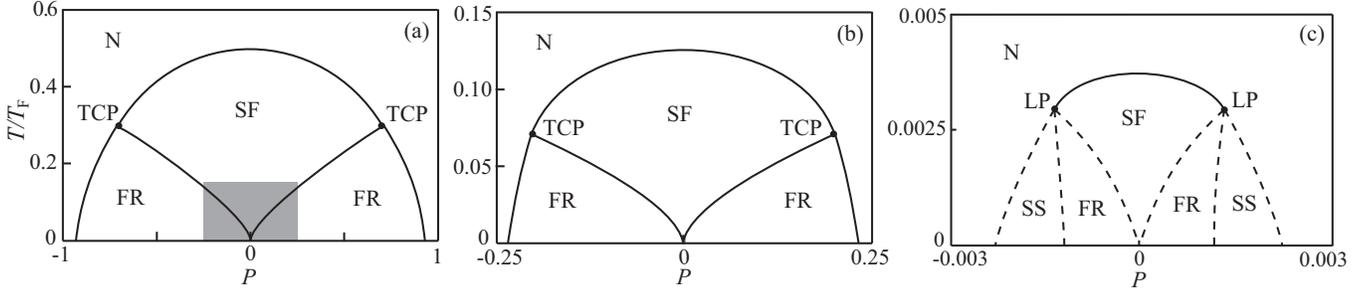}
\caption{Phase diagrams of the mass-balanced Fermi gas, $r=1$, as
a function of temperature $T$ and polarization $P$ at different
interaction strengths. The temperature is scaled with the Fermi
temperature $T_{\text{F}}$. Panel (a) shows the phase diagram in
the strongly interacting limit, $1/k_{\text{F}}a=0$. There is a
tricritical point (TCP), where the normal state (N), the
homogeneous superfluid state (SF) and the forbidden region (FR)
meet. The shaded area sets the scale for panel (b), where the phase
diagram for a weaker interaction, $1/k_{\text{F}}a=-1$, is shown.
Again, there is a tricritical point. There is again a shaded
region to set the scale for panel (c), but it is too small to see.
In panel (c) the phase diagram is shown for a very weak
interaction, $1/k_{\text{F}}a=-3$, and the critical temperatures
are now extremely low. For this interaction we find a Lifshitz
point (LP), below which there is an instability towards a
supersolid (SS). The size of this supersolid region is not
calculated within our theory and therefore the dashed lines are
only guides to the eye.} \label{een}
\end{figure*}

\begin{figure*}
\includegraphics[width=\textwidth]{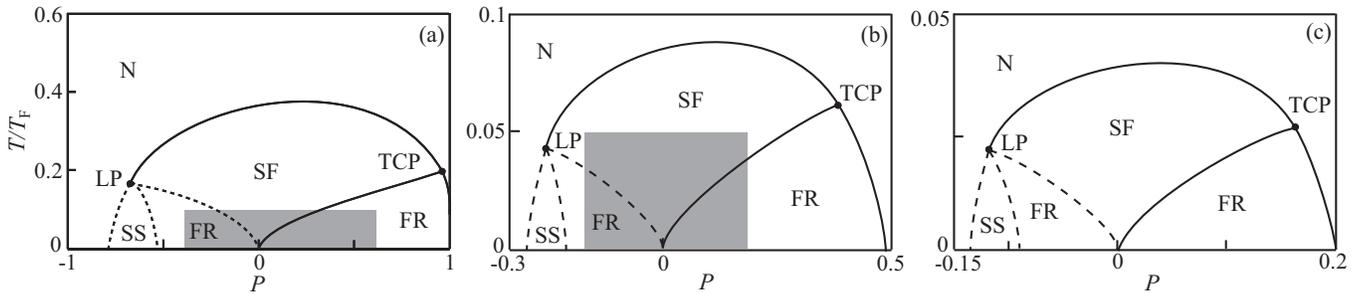}
\caption{Phase diagrams of the Fermi gas with mass ratio $r=6.7$
for different interaction strengths. Panel (a) shows the phase
diagram in the strongly interacting limit. The size of the box is
the same as for the mass-balanced case such that the effect of the
mass imbalance can be seen. The shaded region sets the scale for
panel (b), where the interaction strength is weaker,
$1/k_{\text{F}}a=-1$. And in panel (b) the shaded region in turn
sets the scale for panel (c) where the interaction is even weaker,
$1/k_{\text{F}}a=-3/2$. The dashed lines are again guides to the
eye.}\label{zes}
\end{figure*}

\begin{figure*}
\includegraphics[width=\textwidth]{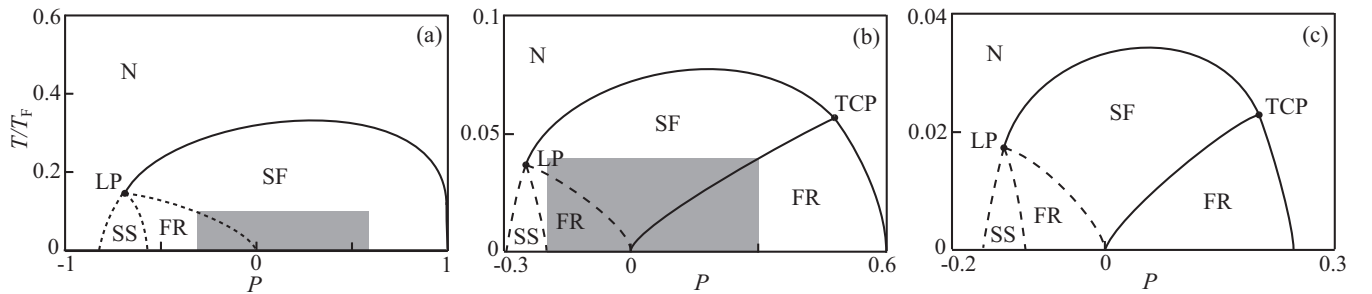}
\caption{Phase diagrams of the Fermi gas with mass ratio $r=10$
for different interaction strengths. Again, panel (a) shows the
phase diagram in the strongly interacting limit and the size of
the box is again the same as for the mass-balanced case. The
shaded region sets the scale for panel (b), where the interaction
strength is weaker, $1/k_{\text{F}}a=-1$. And in panel (b) the
shaded region in turn sets the scale for panel (c) where the
interaction is even weaker, $1/k_{\text{F}}a=-3/2$. Again, the
dashed lines are only guides to the eye.}\label{tien}
\end{figure*}

We present the phase diagram for three different mass ratios,
namely for $r=1$, which is the mass-balanced case, $r=6.7$, which
corresponds to a ${}^{6}$Li-${}^{40}$K mixtures, and for $r=10$
where an interesting feature in the phase diagram is found
regarding the tricritical point \cite{Parish}. For these mass
ratios we present the phase diagrams at three different
interaction strengths in order to see what the effect of the
interaction is on the critical temperature. Namely, we present the
phase diagrams for a strongly interacting Fermi gas with
$1/k_{\text{F}}a=0$, an intermediate interaction strength with
$1/k_{\text{F}}a=-1$ and a weakly interacting Fermi gas. These
interaction strengths correspond to the BCS side of the Feshbach
resonance, see Fig.~\ref{fesh}.

Fig.~\ref{een} shows the phase diagrams of the mass-balanced Fermi
gas at three different interaction strengths. In Fig.~\ref{een}a
the phase diagram for the strongly interacting regime where
$1/k_{\text{F}}a=0$ is depicted. It is symmetric in the
polarizations. The phase transition is of second order for small
polarizations. Moreover, we find two tricritical points in the
phase diagram and below the tricritical points the phase transition is of first order. At a first-order phase transition the order parameter is discontinuous and therefore also the particle densities $n_{\sigma}$, see Eq.~(\ref{particle densities}). And thus also the polarization is discontinuous at a first-order phase transition, which gives rise to a forbidden region where phase separation occurs.
Fig.~\ref{een}b again shows the phase diagram of the mass-balanced
Fermi gas, but now for a weaker interaction, namely for
$1/k_{\text{F}}a=-1$. Compared to the unitarity regime, there are
no qualitative changes in the phase diagram. However,  the
critical temperatures are lower than in the strongly interacting
regime. Then, Fig.~\ref{een}c shows the phase diagram of the
mass-balanced Fermi gas for a very weak interaction, namely
$1/k_{\text{F}}a=-3$. Apart from the fact that the critical
temperatures are now extremely low, there is a large difference
compared to the strongly interacting regime, namely we do not find
tricritical points in the phase diagram but instead we find two
Lifshitz points. Below the Lifshitz point there is an instability
towards supersolidity. We assumed that there will be a second-order transition from
the normal state to the supersolid. The critical temperature for
this transition is found by solving $\alpha(\mathbf{k}_{\text{LO}})=0$, where $\mathbf{k}_{\text{LO}}$ is the wavevector of the supersolid. The transition from supersolidity to the homogeneous superfluid phase is expected
to be first order \cite{Casalbuoni,Mora} and thus a forbidden region will be present in the phase diagram. This simple scenario is sketched in the phase diagram, but it could be that the normal to supersolid second-order phase transition is preempted by weak first-order transitions from the
normal state to various more complicated supersolid phases
\cite{Casalbuoni,Mora,Yip}. For the calculation of the relevant fourth-order diagrams with nonzero external momenta, one needs to make an assumption about the crystal structure of the supersolid phase, in order to calculate a first-order phase transition to the supersolid phase. This calculation for the
stability regions of all possible supersolid phases is beyond the
scope of this paper. To emphasize that we did not investigate this region in great detail we there used dashed lines in the phase diagram.

Fig.~\ref{zes} again shows three phase diagrams at different
interactions, but now for the mass-imbalanced Fermi gas with mass
ratio $r=6.7$, corresponding to the ${}^{6}$Li-${}^{40}$K Fermi
mixture. In the unitarity regime, see Fig.~\ref{zes}a, the phase
diagram is no longer symmetric in polarizations and the
temperatures are lower in comparison with the mass-balanced case.
Furthermore, already in the strongly interacting regime we find a
Lifshitz point in the phase diagram for a majority of heavy atoms.
For a majority of light particles a tricritical point is found.
This is in sharp contrast with the mass-balanced case where a
Lifshitz point is only present for extremely weak interactions. As
a result, the Lifshitz temperature is then at least about a
hundred times lower than for the ${}^{6}$Li-${}^{40}$K mixture.
Below the tricritical point there is a forbidden region and below
the Lifshitz point there is an instability towards a supersolid.
We sketched the same scenario below the Lifshitz point here as for
the mass-balanced case. In Fig.~\ref{zes}b and Fig.~\ref{zes}c the
phase diagrams of the Fermi gas are depicted at interaction
strengths $1/k_{\text{F}}a=-1$ and $1/k_{\text{F}}a=-3/2$
respectively. The interaction strength does not affect the topology
of the phase diagram, but it does affect the critical
temperatures. Just as in the mass-balanced case, the critical
temperatures are lower for weaker interactions, as expected.

In Fig.~\ref{tien} we show the phase diagrams of an imbalanced
Fermi gas with an even larger mass imbalance, namely with mass
ratio $r=10$. The diagram has become even more asymmetric. In the
strongly interacting regime, we also find a Lifshitz point for a
majority of heavy atoms. But the tricritical point has
disappeared, which means that the phase transition remains of
second order and no phase separation occurs for a majority of
light particles. Below the Lifshitz point the same scenario is
sketched as before. Fig.~\ref{tien}b shows the phase diagram at a
weaker interaction strength, namely at $1/k_{\text{F}}a=-1$. With
respect to the unitarity limit there is a real change in this
diagram. Namely, the tricritical point reappears in the phase
diagram. Then, in Fig.~\ref{tien}c the interaction strength is
even weaker, $1/k_{\text{F}}a=-3/2$, and the critical temperatures
are lower. But there are no qualitative changes with respect to
the phase diagram in Fig.~\ref{tien}b.

\subsubsection{Effect of a Mass Imbalance}

By comparing the phase diagrams of the mass-balanced Fermi gas and
of the Fermi gases with a mass imbalance, it can be seen that a
mass imbalance causes important changes. First, the critical
temperatures for the mass-imbalanced Fermi gases are lower than
for the mass-balanced case. Second, for a mass-imbalanced Fermi
gas the phase diagram is no longer symmetric in polarizations. In
other words, the maximum critical temperature is no longer located
at zero polarization. Third, for Fermi gases with a sufficiently
large mass ratio, like the two mass-imbalanced Fermi gases we
presented here, there is a Lifshitz point in the phase diagram for
a majority of heavy atoms. These three changes we now discuss in
some more detail.

\begin{figure}
\includegraphics[width=\columnwidth]{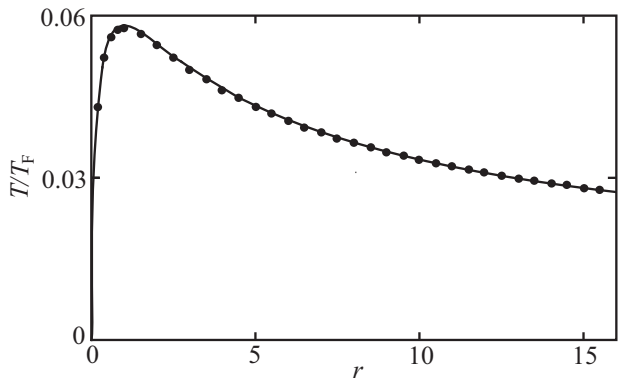}
\caption{The critical temperature of an unpolarized Fermi gas as a
function of the mass ratio $r=m_{-}/m_{+}$. The circles are the
numerical results of $\alpha=0$ and the full line is the analytic
result for the weakly interacting limit. Here
$1/k_{\text{F}}a=-3/2$.}\label{tcvanr}
\end{figure}

In the weakly interacting limit, the critical temperatures are
very low. For the unpolarized Fermi gas the integral in
Eq.~(\ref{alpha}) can then be evaluated exactly. Then, by equating
$\alpha$ to zero an analytic result for the critical temperature
can be obtained \cite{Paananen}
\begin{align}
\nonumber \left. T_{c}\right|_{P=0}&=\sqrt{\varepsilon_{\text{F},+}\varepsilon_{\text{F},-}}\frac{8}{\pi k_{\text{B}}}e^{\gamma-2}e^{-\pi/2k_{\text{F}}|a|}\\
&=\frac{2\sqrt{r}}{1+r}\frac{8\varepsilon_{\text{F}}}{\pi
k_{\text{B}}}e^{\gamma-2}e^{-\pi/2k_{\text{F}}|a|}, \label{tcr}
\end{align}
where $\varepsilon_{\text{F}}$ is the Fermi energy corresponding
to twice the reduced mass, $k_{\text{B}}$ is Boltzmann's constant
and $\gamma=0.5772$ is Euler's constant. When the atoms making up
the Fermi mixture have different masses, $r$ is not equal to one.
In that case the term $2\sqrt{r}/(1+r)$ is smaller than one. And
thus, by adding a mass imbalance the critical temperature for the
unpolarized Fermi gas becomes lower. In Fig.~\ref{tcvanr}, it can
be seen that the dependence of the critical temperature on the
mass ratio is indeed as given by Eq.~(\ref{tcr}) and that the
agreement with the numerical results is very good. For the
polarized Fermi gas, we expect a similar dependence of the
critical temperature on the mass ratio.

\begin{figure}
\includegraphics[width=0.97\columnwidth]{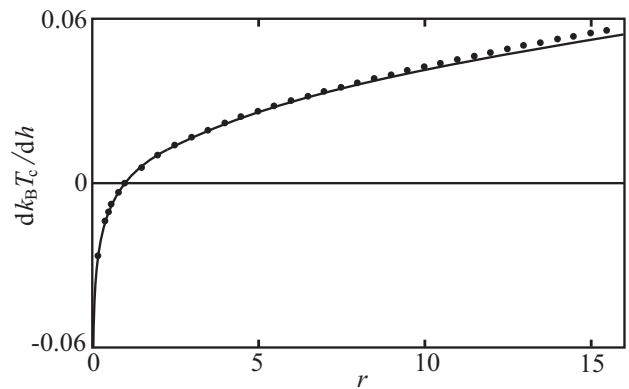}
\caption{Derivative of the critical temperature with respect to
the difference in chemical potentials $h$. The circles are the
numerical results and the full line is the analytic result. The
interaction strength is $1/k_{\text{F}}a=-3/2$.}\label{afgeleide}
\end{figure}

For the mass-balanced Fermi gas, the highest critical temperature
is located at zero polarization, see Fig.~\ref{een}, whereas for
the mass-imbalanced Fermi gas it is located at nonzero
polarization, see Fig.~\ref{zes} and Fig.~\ref{tien}. As a
consequence, the derivative of the critical temperature with
respect to $h$, the difference in chemical potentials, will no
longer be zero at $P=0$. In the weakly interacting limit, the
coefficient $\alpha$ in Eq.~(\ref{alpha}) can be expanded
around the critical temperature $T_{c}$ and the critical `Zeeman'
field $h_{0}$ corresponding to the unpolarized Fermi gas
\begin{align}
\nonumber&\alpha(T_{c}+\delta T_{c},h_{0}+\delta h)\simeq\\
&\alpha(T_{c},h_{0})+\frac{\partial\alpha}{\partial
T}(T_{c},h_{0})\delta T_{c}+\frac{\partial\alpha}{\partial
h}(T_{c},h_{0})\delta h.
\end{align}
For a continuous phase transition, $\alpha$ has to go to zero. The
term on the left-hand side and the first term on the right-hand
side are therefore zero. Then we find from the other two terms
\begin{align}
\nonumber \left.\frac{\delta T_{c}}{\delta h} \right|_{P=0}&=-\frac{\partial\alpha/\partial h}{\partial\alpha/\partial T}=\left.T_{c}\right|_{P=0}\frac{1}{4}\left(\frac{1}{\varepsilon_{\text{F},-}}-\frac{1}{\varepsilon_{\text{F},+}}\right)\\
&=\frac{2}{\pi
k_{\text{B}}}\frac{r-1}{\sqrt{r}}e^{\gamma-2}e^{-\pi/2k_{\text{F}}|a|}.
\label{derivtc}
\end{align}
This is indeed nonzero if the mass ratio is not equal to one.
Eq.~(\ref{derivtc}) is positive for mass ratios larger than one.
This is in agreement with our findings that the phase diagram
shifts towards positive polarizations for a mass ratio larger than
one, see Fig.~\ref{zes} and Fig.~\ref{tien}. For mass ratios
smaller than one, Eq.~(\ref{derivtc}) is negative and the phase
diagram then shifts towards negative polarizations. In both cases
the phase diagram shifts towards a majority of light particles.
Eq.~(\ref{derivtc}) is plotted in Fig.~\ref{afgeleide}, together
with the numerical results. The agreement is good, especially for
small mass ratios.

\begin{figure}
\includegraphics[width=.9\columnwidth]{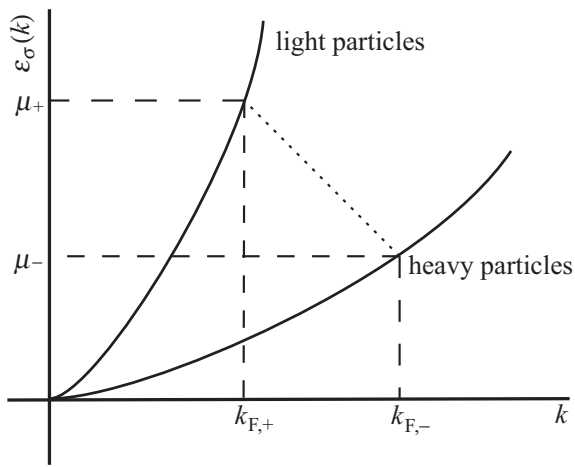}
\caption{The energies $\varepsilon_{\sigma}(k)$ of the fermions
for a Fermi mixture with both a population and a mass imbalance,
where $m_{-}>m_{+}$ and $n_{-}>n_{+}$. Thus, there is a majority
of heavy particles. At the Fermi levels, the energy difference
between particles with different momenta (dashed line) is smaller
than the energy difference between particles with the same
momenta.} \label{physics}
\end{figure}
For a sufficiently large mass ratio we find a Lifshitz point in
the phase diagram for a majority of heavy particles. To explain
why this can be expected we consider the kinetic energies of the
particles in Eq.~(\ref{energies}), which are plotted in
Fig.~\ref{physics}. In the mass-balanced case a fermion of one
species with momentum $\mathbf{k}$ has the same energy as a
fermion with momentum $-\mathbf{k}$ of the other species. Coupling
these degenerate states by a condensate of Cooper pairs is
energetically desirable since the shift of the energy levels due
to the coupling is now largest, as is well-known from the physics
of avoided crossings. Therefore, it is thus most favorable to form
a pair of two fermions with the same but opposite momentum, i.e.,
a fermion with momentum $\mathbf{k}$ forms a pair with a fermion
with momentum $-\mathbf{k}$, since then the energy gain as a
result of pairing is maximal. The pair then has no kinetic energy
and the pairs form a homogeneous superfluid.

Typically, the formation of pairs mostly occurs at the Fermi energy. There a pair
consists of one fermion with momentum $\mathbf{k}_{\text{F}}$ and
one with momentum $-\mathbf{k}_{\text{F}}$. Since the Fermi
momentum depends on the density of the particles $n_{\sigma}$, see
Eq.~(\ref{fermimomentum}), the Fermi momenta will not be the same
in the case of a population imbalance. This makes pairing less
ideal than in the unpolarized case. The critical temperatures are
therefore lower for a polarized mixture, see Fig.~\ref{een}.

In the mass-imbalanced case two fermions with equal momentum but
from different species do not have the same kinetic energy. Thus, if now a
fermion of one species with momentum $\mathbf{k}$ forms a pair
with a fermion with momentum $-\mathbf{k}$ of the other species,
there is an energy difference that reduces the energy gain that
can be obtained due to pairing. This explains the fact that the
critical temperatures are lower for Fermi mixtures with a mass
imbalance. With both a mass imbalance and a population imbalance
pairs can still be formed with two fermions with equal but
opposite momentum, even though there is an energy difference
between the Fermi levels. However, this occurs only if the mass
ratio and the population imbalance are not too large. In the
extreme case of a large mass imbalance and a large population
imbalance a more favorable scenario is possible, see
Fig.~\ref{physics}. In this figure the kinetic energies are
plotted for the different fermions, and there is a majority of
heavy particles, thus $k_{\text{F},-}>k_{\text{F},+}$. It can be
seen that the energy difference between particles with different
momenta is now smaller than the energy difference between fermions
with the same momentum, i.e., it is now energetically more
favorable to form pairs with a nonzero momentum that allows for a
direct coupling of the Fermi levels. The first point where this is
the case is the Lifshitz point. There the system becomes a
supersolid.

\subsubsection{Multicritical Point}
\begin{figure*}
\includegraphics[width=\textwidth]{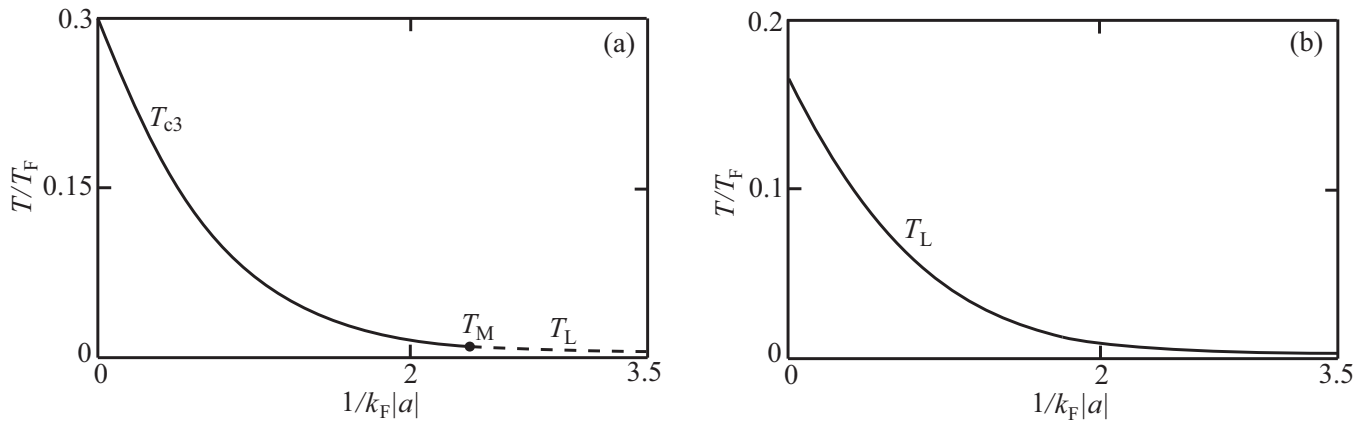}
\caption{Temperature of the tricritical points and the Lifshitz
points as a function of the interaction strength. Panel (a) shows
the mass-balanced case. In panel (b) we look at a Fermi gas with
mass ratio $r=6.7$ and a majority of heavy particles.
$T_{\text{c}3}$ is the temperature of a tricritical point,
$T_{\text{M}}$ is the temperature of a multicritical point and
$T_{\text{L}}$ of a Lifshitz point.} \label{tc3}
\end{figure*}
In the phase diagrams, Fig.~\ref{een}, Fig.~\ref{zes} and
Fig.~\ref{tien}, we find, depending on the mass ratio and the
interaction strength, tricritical points $(P_{\text{c}3},T_{\text{c}3})$, which are solutions of Eq.~(\ref{tricrit}), as well as Lifshitz points
$(P_{\text{L}},T_{\text{L}})$, solutions of Eq.~(\ref{lifshitz}).

At unitarity, we find for the mass-balanced case at unitarity a tricritical point, whereas we find a Lifshitz point for the ${}^{6}$Li-${}^{40}$K mixture for a majority of heavy particles. This Lifshitz point is a clearly distinct point from a tricritical point, just as the tricritical point for the mass-balanced case is a clearly distinct point from a Lifshitz point. This is in accordance with the findings of Parish {\it et al.} \cite{Marchetti}, who found that for a mass-balanced Fermi mixture the FFLO line detaches from the tricritical point away from the BCS limit.
However, there is a mass ratio between $r=1$ and $r=6.7$ for which the phase diagram at unitarity contains a point which is both a tricritical point and a Lifshitz point. In other words, for this mass ratio Eq.~(\ref{tricrit}) and Eq.~(\ref{lifshitz}) have a solution exactly at the same point in the phase diagram. This is a multicritical point and it is found to occur for the mass ratio $r=4.22$.

It can also be seen in the different phase
diagrams that for a given mass ratio the location of tricritical and Lifshitz points
changes as the interaction strength changes. In Fig.~\ref{tc3}, it
is shown for the mass-balanced Fermi gas and for the Fermi gas
with mass ratio $r=6.7$ how the temperature $T$ of these points
changes as a function of the interaction strength. Note that the
polarization is not constant for the lines in Fig.~\ref{tc3}. In
both Fig.~\ref{tc3}a and Fig.~\ref{tc3}b one can recognize an
exponential decay as a function of the interaction strength, which
is exactly what one would expect from Eq.~(\ref{tcr}).

Fig.~\ref{tc3}a shows the mass-balanced case. In the unitarity
regime and for small values of $1/k_{\text{F}}|a|$ a tricritical
point is present in the phase diagram, with temperature
$T_{\text{c}3}$. This is the full line in Fig.~\ref{tc3}a. Then
for some value of the interaction strength Eq.~(\ref{tricrit}) and
Eq.(\ref{lifshitz}) have a solution at the same point. Thus, there
is a multicritical point and the temperature
of this point is in Fig.~\ref{tc3} denoted by $T_{\text{M}}$. For
even weaker interactions it then turns out that there is an
instability towards a supersolid and thus we find a Lifshitz point
in the phase diagram, with temperature $T_{\text{L}}$. This is the
dashed line in Fig.~\ref{tc3}a.

Although the tricritical and Lifshitz points are very close together for very weak interactions, in our approach, where we perform the momentum integrals to calculate the coefficients $\gamma$ in Eq.~(\ref{gamma}) and $\beta$ in Eq.~(\ref{beta}), they are essentially always two distinct points. Except for one value of the interaction strength where we find a multicritical point. If one, however, assumes particle-hole symmetry in the weakly-interacting limit, Eq.~(\ref{gamma}) and Eq.~(\ref{beta}) become, up to a constant, the same equation and thus one finds that the tricritical point and the Lifshitz point are the same point \cite{Casalbuoni,Buzdin}.

Fig.~\ref{tc3}b gives the temperature of the Lifshitz point
present in the phase diagram of the Fermi gas with mass ratio
$r=6.7$ for a majority of heavy particles. Here, we do not find a
change in character of the phase transition as a function of
interaction strength. The Lifshitz point remains a Lifshitz point.
For a majority of light particles of this Fermi mixture, we found
a tricritical point in the strongly interacting limit. We expect
that this point will also change to a Lifshitz point for very weak
interactions, but the extremely low temperatures make it
numerically difficult to investigate this possibility in detail.
\subsubsection{Sarma Phase}
\begin{figure*}[t]
\includegraphics[width=\textwidth]{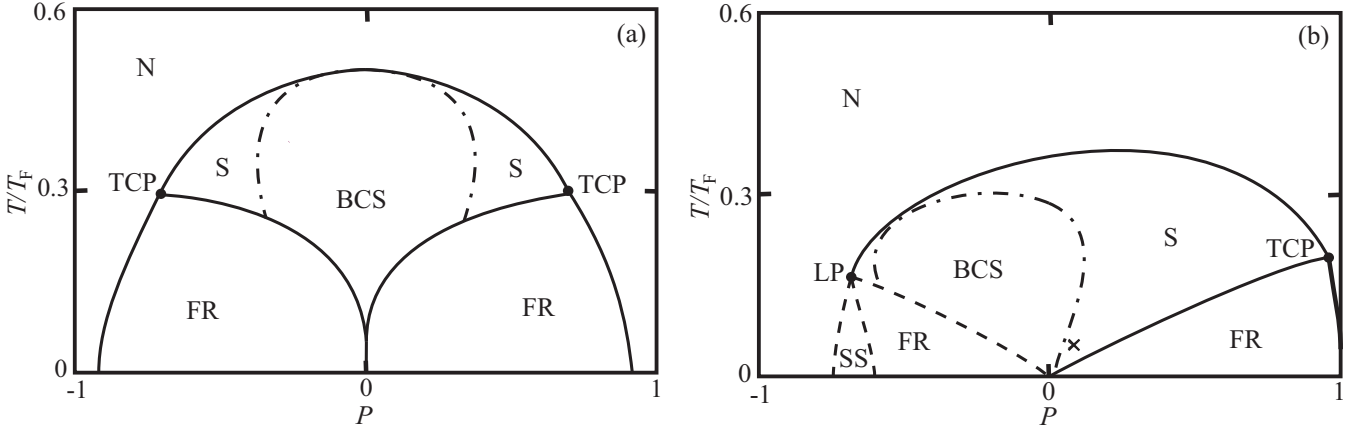}
\caption{Phase diagrams at unitarity including the region where
the superfluid is gapless. Panel (a) shows the mass-balanced case.
It is the same phase diagram as in Fig.~8a, only now the
superfluid region is divided into a gapped superfluid region (BCS)
and two Sarma regions (S). Panel (b) shows the mass-imbalanced
Fermi gas with mass ratio $r=6.7$. The small cross in panel (b) is
the point at which we calculated the distribution functions of the
particles shown in Fig.~\ref{distributiessarma}.}
\label{sarmaplaatje}
\end{figure*}

In Sec.~IVA, while explaining the dispersions of the
quasiparticles, we discussed the possibility of a Sarma phase,
which occurs when one of the dispersions
$\hbar\omega_{\sigma}(\mathbf{k})$ becomes negative. At each point
in the phase diagram the value of the order parameter $|\Delta|$
can be found by minimizing the thermodynamic potential in
Eq.~(\ref{thermpot}). With this value of the order parameter it
can be determined whether the dispersions are always positive or
become negative for some range of momenta. In the superfluid phase
it turns out that there are different regions. In one region the
dispersions are always positive and the superfluid is gapped,
while in the other region one of the dispersions becomes negative
and we are dealing with a gapless superfluid. At nonzero
temperatures, there is no phase transition between these regions.
Rather, there is only a smooth crossover. We calculated the
gapless superfluid region in the unitarity limit for the
mass-balanced case and for the ${}^{6}$Li-${}^{40}$K Fermi
mixture. The results are shown in Fig.~\ref{sarmaplaatje}. In the
mass-balanced case, Fig.~\ref{sarmaplaatje}a, there are two
regions in the phase diagram where we have a gapless superfluid.
In Fig.~\ref{sarmaplaatje}b the mass-imbalanced case is shown.
Here, the Sarma region is not symmetric in polarizations, just as
the rest of the phase diagram. The Sarma region is very large for
positive polarizations, whereas for negative polarizations it
becomes very small.

\begin{figure}
\includegraphics[width=\columnwidth]{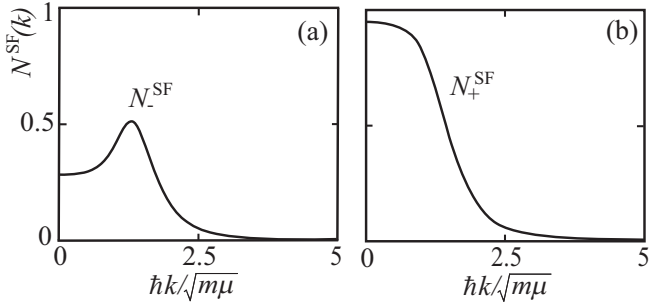}
\caption{The distribution functions
$N_{\sigma}^{\text{SF}}(\mathbf{k})$ for the heavy particles,
panel (a), and the light particles, panel (b), in the gapless
superfluid phase.} \label{distributiessarma}
\end{figure}

Fig.~\ref{distributiessarma} depicts the distribution functions
$N_{\sigma}^{\text{SF}}(\mathbf{k})$ of the heavy and the light
particles for some point in the phase diagram of
Fig.\ref{sarmaplaatje}b that lies in the superfluid Sarma region.
In the superfluid phase the distribution functions are modified
with respect to the Fermi distribution functions
$N_{\sigma}(\mathbf{k})$ and are given by
\begin{align}
\nonumber N_{\sigma}^{\text{SF}}(\mathbf{k})&=\frac{1}{2}\left(1+\frac{\varepsilon(\mathbf{k})-\mu}{\hbar\omega(\mathbf{k})}\right)N_{\sigma}(\mathbf{k})\\
&+\frac{1}{2}\left(1-\frac{\varepsilon(\mathbf{k})-\mu}{\hbar\omega(\mathbf{k})}\right)(1-N_{-\sigma}(\mathbf{k})).
\end{align}
It can be seen in Fig.~\ref{distributiessarma} that the
distribution function of the heavy particles is nonmonotic, which
is a signature of the Sarma phase.

\section{Fluctuation Effects}

\begin{figure*}
\includegraphics[width=0.85\textwidth]{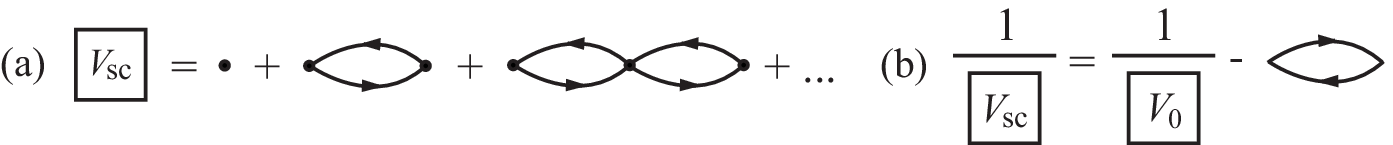}
\caption{The screened interaction containing an infinite sum of
bubble diagrams. Panel (a) shows the perturbative expansion of the screened interaction and panel (b) shows the result of the resummation of this expansion.} \label{bubble}
\end{figure*}
\begin{figure*}
\includegraphics[width=0.85\textwidth]{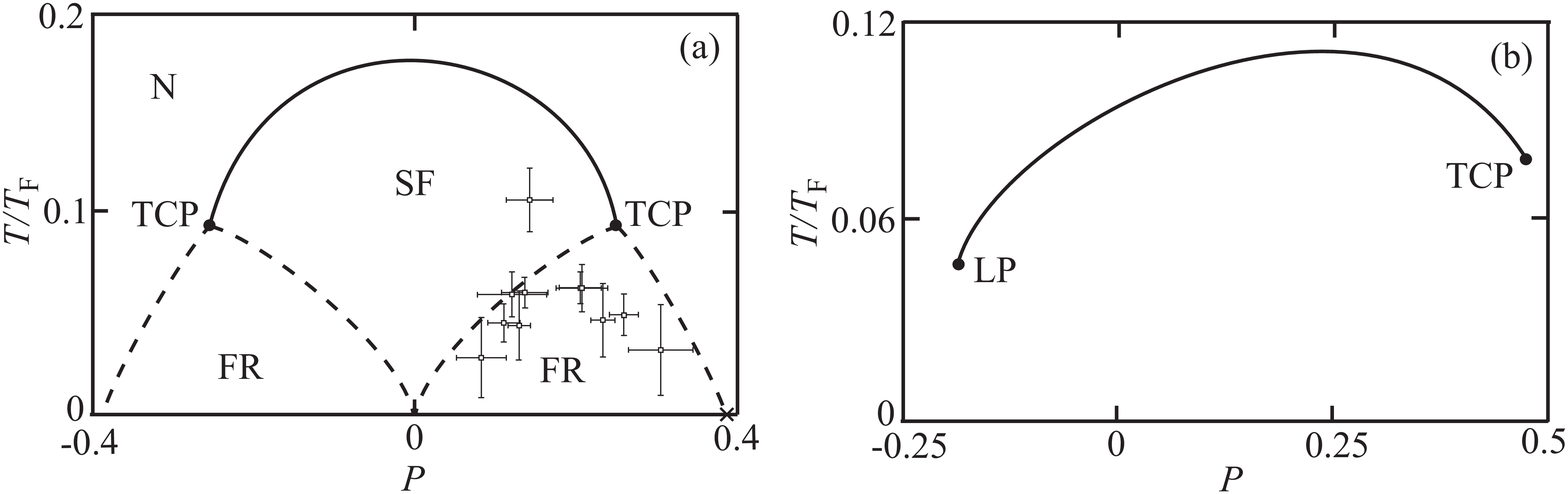}
\caption{Phase diagrams of strongly interacting Fermi gases, i.e.,
$ 1/k_{\text{F}}a=0$, where fluctuation effects have been taken
into account. Panel (a) shows the mass-balanced case,i.e., $r=1$. The open
squares are data along the phase boundaries from the experiment of
Shin {\it et al}. \cite{Shin}. The dashed lines denote the first-order
phase transitions, not calculated within our theory and therefore only guides to the eye. However, they include the correct critical polarization at zero temperature, denoted by a cross, known from Monte Carlo calculations \cite{Lobo}. To calculate the first-order phase transition lines
more accurately with fluctuation effects taken into account other
methods are needed \cite{Diederix}. Panel (b) shows the phase
diagram for the ${}^6$Li-${}^{40}$K mixture, where $r=6.7$. For both mixtures the
critical temperatures are lowered by taking fluctuation effects
into account, compare Fig.~\ref{een}a and Fig.~\ref{zes}a.}
\label{r1 met fluctuaties}
\end{figure*}
As argued before, mean-field theory only leads to a qualitative
description of the phase transitions that occur in an imbalanced
Fermi gas with unitarity-limited interactions. To achieve also a
quantitative description, we have to take fluctuation effects into
account. From renormalization-group calculations \cite{Koos}, we
know that especially selfenergy effects and the screening of the
interaction by particle-hole fluctuations are important in the
unitarity limit. The corresponding corrections result in
quantitative agreement with experiments in the mass-balanced case
\cite{Koos}, and give rise to more accurate predictions for
upcoming experiments with the very promising ${}^{6}$Li-${}^{40}$K
mixture \cite{prl}.

As explained in Refs. \cite{Diederix,prl}, a convenient way to
take selfenergy corrections into account is to introduce
renormalized chemical potentials $\mu'_{\sigma}$ that describe the
selfenergy of particles with spin $\sigma$ in a Fermi sea of
particles with spin $-\sigma$. This can be achieved by using
\begin{eqnarray}
\mu'_{\sigma}=
\mu_{\sigma}+c_{\sigma}\frac{\mu'^2_{-\sigma}}{\mu'_{\sigma}+\mu'_{-\sigma}}
\label{Selfenergy2},
\end{eqnarray}
where $c_{\sigma}$ is a coefficient that can be determined in the
ladder approximation \cite{Combescot}, but also with the use of
renormalization-group calculations \cite{Koos} or Monte Carlo
calculations \cite{Carlson}. For the mass-balanced case, we use
$c_{\pm}=0.6$, while for the ${}^{6}$Li-${}^{40}$K mixture we use
$c_{+}=2.3$ and $c_{-}=0.36$ to incorporate the Monte Carlo
results. The substitution of these renormalized chemical
potentials $\mu'_{\sigma}(\mu_{\sigma})$ in the mean-field
thermodynamic potential $\omega_{\text{L}}(\Delta;\mu'_{\sigma})$
results in the following equation for the densities
\begin{equation}
n_{\sigma}=-\left.\frac{\partial\omega_{\text{L}}(\Delta;\mu'_{\sigma})}{\partial\mu_{\sigma}}\right|_{\Delta=\langle\Delta\rangle}.
\end{equation}
In Ref. \cite{prl}, an accurate comparison between the use of
renormalized chemical potentials and Monte Carlo calculations was
made at zero temperatures, leading to excellent agreement. Note that in the superfluid state Eq.~(\ref{Selfenergy2}) overestimates the effects of the interactions on the renormalization of the chemical potentials and another correction proportional to $\Delta^2$ has to be subtracted in the right-hand side of Eq.~(\ref{Selfenergy2}) \cite{Diederix}.

Fluctuations do not only affect the selfenergies of the fermions
in the normal state. There is another effect of particle-hole
fluctuations that affects the transition to the superfluid state. Namely, there is a
change in the coefficient $\alpha$ due to screening of the
interspecies interaction, which is also called the Gor'kov
correction. We can take the screening into account by considering
an effective two-body interaction that includes the so-called random-phase approximation (RPA)
bubble sum. This procedure is diagrammatically represented in
Fig.\ \ref{bubble}. The inclusion of the infinite geometric series
of bubble diagrams leads at zero external momentum and frequency
to
\begin{equation}\label{vsc}
\frac{1}{V_{\rm sc}}=\frac{1}{V_0}-\hbar\Pi({\bf 0},0),
\end{equation}
where $\hbar\Pi({\bf 0},0)$ is the amplitude of the bubble
diagram. In Appendix B, it is shown that this amplitude is given
by
\begin{eqnarray}\label{Ladder2}
\hbar \Pi({\bf 0},0)&=&\int \frac{d{\bf k' }} {(2\pi)^3}
\frac{N_+({\bf k'})-N_-({\bf k'})}{2h' - \epsilon_{+}({\bf
k'})+\epsilon_{-}({\bf k'})},
\end{eqnarray}
where we use renormalized chemical potentials to also include the
fermionic selfenergy effects. Using the screened interaction of
Eq. (\ref{vsc}) in Eq. (\ref{t2b bij puntinteractie}), we obtain a
screened two-body transition matrix $T^{\rm 2B}_{\rm sc}(0)$,
which consequently enters the expression for the quadratic
coefficient in Eq. (\ref{alpha}). When the quadratic coefficient
including screening becomes zero, a second-order transition can
occur, so that the critical condition now becomes
\begin{equation}
\alpha_{\rm sc}(T_{\rm c},\mu_{\sigma}')=\alpha(T_{\rm
c},\mu_{\sigma}')+\hbar\Pi({\bf 0},0)=0.
\end{equation}
The new critical condition that includes both screening and
fermionic selfenergy effects typically reduces the obtained
critical temperatures with a factor of three.

If we apply this procedure to determine the line of second-order
phase transitions in the mass-balanced case, we obtain the result
in Fig.\ \ref{r1 met fluctuaties}a. In this figure, also the data
along the phase boundaries from the experiment of Shin {\it et
al.} \cite{Shin} are shown. For the unpolarized Fermi gas, we find
$T_{\rm c} = 0.18 T_{\rm F}$ and $\mu(T_{\rm c})=0.51 T_{\rm F}$
\cite{prl}, which is to be compared with the Monte Carlo results
$T_{\rm c} = 0.15 T_{\rm F}$ and $\mu(T_c)=0.49 T_{\rm F}$
\cite{Burovski}. Moreover, for the location of the mass-balanced
tricritical point we find $k_{\rm B}T_{{\rm c}3} = 0.09
\varepsilon_{\rm F+}$ and $P_{{\rm c}3}=0.24$ \cite{prl}, which is
rather close to the experimental data \cite{Shin}. We included the critical polarization at zero temperature known from Monte Carlo calculations \cite{Lobo}, in order to be able to sketch the first-order line and the forbidden region. Other methods are needed to calculate the first-order phase transition more accurately with fluctuation effects taken into account \cite{Diederix}. In Fig. \ref{r1 met fluctuaties}b, we show the line of second-order phase
transitions for the ${}^6$Li-${}^{40}$K mixture, when fluctuation
effects are included. Compared to the mean-field result of Fig.\
\ref{sarmaplaatje}b, the critical temperatures are significantly
lower. For the positions of the tricritical point and the Lifshitz
point, we use Eq. (\ref{beta}) and Eq. (\ref{gamma}), where we
again insert the renormalized chemical potentials to include the
fermionic selfenergy effects. We then find $k_{\rm B}T_{{\rm c}3}
= 0.08T_{\text{F}}$ and $P_{{\rm c}3}=0.47$, and $k_{\rm
B}T_{\text{L}} = 0.05T_{\text{F}} $ and $P_{\text{L}}=-0.18$,
respectively \cite{prl}. It is important to note that although the
fluctuations have quantitatively a very large effect, the topology
of the phase diagrams remains the same.

\section{Discussion and Conclusion}

In this paper we considered Fermi mixtures consisting of two
different species of fermions, which can both have a population
and a mass imbalance. On the mean-field level we calculated the
phase diagrams for those Fermi mixtures as a function of
temperature and polarization. We calculated the phase diagrams for
different mass ratios and for different interaction strengths,
where we found that a mass imbalance leads to a phase diagram that
is asymmetrical in the polarization. We also considered the
possibility of a Lifshitz instability. We found such instabilities
in the Fermi mixtures with a mass ration of $r=6.7$ and $r=10$ in
the unitarity limit. For a mass-balanced mixture, Lifshitz points
occur only in the weakly interacting regime.

By studying the effects of a mass imbalance in more detail we
found analytic results for the critical temperature and the change
in critical temperature at zero polarization. Furthermore, we
investigated what happens to the position of the Lifshitz points
and the tricritical points when changing the interaction strength.
In the mass-balanced case, we also found a multicritical point for
weak interactions. Both for the mass-balanced and the
${}^6$Li-${}^{40}$K Fermi mixtures, we calculated the regions
where the superfluid phase is gapless in the unitarity limit,
i.e., where the mixtures are in the Sarma phase. These regions
were present at nonzero temperatures, and turned out to be quite
large.

Finally, to obtain more quantitative results we introduced
renormalized chemical potentials that include selfenergy effects
to account for the resonant interactions. We also took screening
effects on the critical temperature into account. In this way, we obtained a phase diagram for the mass-balanced mixture that agrees well with Monte Carlo calculations and with experiment. And we hope to have obtained a good quantitative description of the phase
diagram for the ${}^6$Li-${}^{40}$K mixture, where especially the
presence of a Lifshitz point is exciting. Below the Lifshitz point
various supersolid states are competitive. This leads to open
questions for further research on the interesting possibility of
an inhomogeneous superfluid.

\section{Acknowledgements}
This work is supported by the Stichting voor Fundamenteel
Onderzoek der Materie (FOM) and the Nederlandse Organisatie voor
Wetenschaplijk Onderzoek (NWO).

\appendix
\section{Derivation of the Mean-Field Thermodynamic Potential}

In this Appendix we explicitly derive the mean-field thermodynamic
potential in Eq.~(\ref{thermpot}). We start with the microscopic
action for an interacting Fermi mixture consisting of fermions
present in two hyperfine states. It is given by
\begin{align}
\nonumber &S[\phi^{*},\phi]=\int_{0}^{\hbar\beta}\text{d}\tau\int\text{d}\mathbf{x}\Big\{\\
\nonumber&\times\sum_{\sigma=\pm}\phi_{\sigma}^{*}(\mathbf{x},\tau)\left(\hbar\frac{\partial}{\partial\tau}-\frac{\hbar^{2}\nabla^{2}}{2m_{\sigma}}-\mu_{\sigma}\right)\phi_{\sigma}(\mathbf{x},\tau)\\
&+V_{0}\phi_{+}^{*}(\mathbf{x},\tau)\phi_{-}^{*}(\mathbf{x},\tau)\phi_{-}(\mathbf{x},\tau)\phi_{+}(\mathbf{x},\tau)\Big\},
\label{microscopic action}
\end{align}
where $V_{0}\,\delta(\mathbf{x-x'})$ is the bare point-like interaction associated with the short-range atomic potentials. The last term in this action is a fourth-order term in the fermionic fields. Therefore, the integral over the fermionic fields in the partition function
\begin{eqnarray}
Z=\int\text{d}[\phi^{*}]\text{d}[\phi]\exp\{-S[\phi^{*},\phi]/\hbar\}
\label{partition function}
\end{eqnarray}
cannot be performed analytically. To deal with this fourth-order
term we introduce by means of a Hubbard-Stratonovich
transformation bosonic pairing fields $\Delta(\mathbf{x},\tau)$,
which are on average related to the fermionic fields
$\phi_{\sigma}(\mathbf{x},\tau)$ as in  Eq.~(\ref{bosonic
fields}). The Hubbard- Stratonovich transformation is performed by
inserting into the partition function the following identity
\begin{equation}
1=\int\text{d}[\Delta^{*}]\text{d}[\Delta]e^{(\Delta-\phi_{+}\phi_{-}V_{0}|V_{0}^{-1}|\Delta^{*}-V_{0}\phi^{*}_{-}\phi_{+}^{*})/\hbar},
\label{hubbard}
\end{equation}
where the inner product in the exponent is a short hand notation
for
\begin{align}
\nonumber\noindent\int_{0}^{\hbar\beta}\text{d}\tau\int\text{d}\mathbf{x}&(\Delta^{*}(\mathbf{x},\tau)-\phi^{*}_{+}(\mathbf{x},\tau)\phi_{-}^{*}(\mathbf{x},\tau)V_{0})\\
&V_{0}^{-1}(\Delta(\mathbf{x},\tau)-V_{0}\phi_{-}(\mathbf{x},\tau)\phi_{+}(\mathbf{x},\tau)).
\end{align}

Inserting Eq.~(\ref{hubbard}) into Eq.~(\ref{partition function})
leaves us with an action $S[\Delta^{*},\Delta,\phi^{*},\phi]$
which depends only quadratically on the fermionic fields
$\phi_{\sigma}(\mathbf{x},\tau)$. It is given by
\begin{align}
\nonumber&S[\Delta^{*},\Delta,\phi^{*},\phi]=-\int_{0}^{\hbar\beta}\text{d}\tau\int\text{d}\mathbf{x}\Bigg\{\frac{|\Delta(\mathbf{x},\tau)|^{2}}{V_{0}}\\
\nonumber&-\hbar\sum_{\sigma=\pm}\int_{0}^{\hbar\beta}\text{d}\tau'\int\text{d}\mathbf{x'}\phi^{*}_{\sigma}(\mathbf{x},\tau)G_{0;\sigma}^{-1}(\mathbf{x},\tau;\mathbf{x'},\tau')\phi_{\sigma}(\mathbf{x'},\tau')\\
&+\phi_{+}^{*}(\mathbf{x},\tau)\phi_{-}^{*}(\mathbf{x},\tau)\Delta(\mathbf{x},\tau)+\Delta^{*}(\mathbf{x},\tau)\phi_{-}(\mathbf{x},\tau)\phi_{+}(\mathbf{x},\tau)\Bigg\}.
\label{actie voor diagrammen}
\end{align}
The noninteracting Green's function is given by
\begin{align}
\nonumber&G_{0;\sigma}^{-1}(\mathbf{x},\tau;\mathbf{x'},\tau')\\
&\hspace{2mm}=\frac{1}{\hbar}\left\{\hbar\frac{\partial}{\partial\tau}-\frac{\hbar^{2}\nabla^{2}}{2m_{\sigma}}-\mu_{\sigma}\right\}\delta(\mathbf{x-x'})\delta(\tau-\tau').
\end{align}
Note that by performing the Hubbard-Stratonovich transformation we
have introduced the order parameter $\Delta(\mathbf{x},\tau)$ in
an exact manner into the many-body theory. If we now integrate out
the fermionic fields we will be left with an effective action
$S^{\text{eff}}[\Delta^{*},\Delta]$, which is directly related to
the Landau free energy. If we apply mean-field theory we assume
that the bosonic fields are position and time independent, i.e.,
$\Delta(\mathbf{x},\tau)=\Delta$. In order to integrate out the
fermionic fields, we need to diagonalize the action. First,
Fourier transforming the above action and then writing it in a
more compact way using matrix multiplication yields
\begin{align}
\nonumber&S[\Delta^{*},\Delta,\phi^{*},\phi]=-\hbar\beta V\frac{|\Delta|^{2}}{V_{0}}-\hbar\beta\sum_{\mathbf{k}}(\varepsilon_{-}(\mathbf{k})-\mu_{-})\\
&-\hbar\sum_{\mathbf{k},n}\left[\phi_{n,+}^{*}(\mathbf{k}),\phi_{-n,-}(\mathbf{-k})\right]\mathbf{G}^{-1}_{\Delta}(\mathbf{k},\text{i}\omega_{n})\left[\phi_{n,+}(\mathbf{k})\atop{\phi_{-n,-}^{*}(\mathbf{-k})}\right],
\label{actie na hubbard}
\end{align}
where $V$ is the volume. In the above expression
\begin{align}
\nonumber&-\hbar\mathbf{G}_{\Delta}^{-1}(\mathbf{k},\text{i}\omega_{n})=\\
&\left[
\begin{array}{cc}
-\text{i}\hbar\omega_{n}+\varepsilon_{+}(\mathbf{k})-\mu_{+}&\Delta\\
\Delta^{*}&-(\text{i}\hbar\omega_{n}+\varepsilon_{-}(\mathbf{k})-\mu_{-})\end{array}\right],
\end{align}
where $\omega_{n}$ are the odd Matsubara frequencies. By writing
the action in matrix form, we have interchanged the fermionic
fields $\phi_{-}$ and $\phi_{-}^{*}$ and thereby we have picked up
a constant term, namely the sum
$\Sigma_{\mathbf{k}}(\varepsilon_{-}(\mathbf{k})-\mu_{-})$ in the first
line of Eq.~(\ref{actie na hubbard}). We can now diagonalize this
action by means of a Bogoliubov transformation. This
transformation consists of unitarily transforming the atomic
fields $\phi_{n,\sigma}(\mathbf{k})$  to the quasiparticle fields
$\psi_{n,\sigma}(\mathbf{k})$
\begin{eqnarray}
\left[\psi_{n,+}(\mathbf{k})\atop{\psi_{-n,-}^{*}(\mathbf{-k})}\right]=\left[\begin{array}{cc}u(\mathbf{k})&-v(\mathbf{k})\\v^{*}(\mathbf{k})&u^{*}(\mathbf{k})\end{array}\right]\cdot\left[\phi_{n,+}(\mathbf{k})\atop{\phi_{-n,-}^{*}(\mathbf{-k})}\right],
\end{eqnarray}
where this transformation is unitary if
\begin{equation}
|u(\mathbf{k})|^{2}+|v(\mathbf{k})|^{2}=1.
\end{equation}
The coefficients of this transformation can be found by demanding
the off-diagonal matrix elements to be zero. The action in terms
of the fields $\psi_{n,\sigma}(\mathbf{k})$ then reads
\begin{align}
\nonumber&S[\Delta^{*},\Delta,\psi^{*},\psi]=-\hbar\beta V\frac{|\Delta|^{2}}{V_{0}}\\
\nonumber&-\hbar\beta\sum_{\mathbf{k}}\big(\varepsilon(\mathbf{k})-\mu-\hbar\omega(\mathbf{k})\big)\\
\nonumber&+\sum_{\mathbf{k},n}\left[(-\text{i}\hbar\omega_{n}+\hbar\omega_{+}(\mathbf{k}))\psi_{n,+}^{*}(\mathbf{k})\psi_{n,+}(\mathbf{k})\right.\\
&\left.+(-\text{i}\hbar\omega_{n}+\hbar\omega_{-}(\mathbf{k}))\psi_{n,-}^{*}(\mathbf{k})\psi_{n,-}(\mathbf{k})\right],
\label{feynmanactie}
\end{align}
where the extra term $\hbar\omega_{+}(\mathbf{k})$ inside the first
sum  again comes from interchanging fermionic fields. The
mean-field partition function reads after performing the
Hubbard-Stratonovich transformation and the Bogoliubov
transformation
\begin{align}
\nonumber Z&=\int\text{d}[\psi^{*}]\text{d}[\psi]\exp\{-S[\Delta^{*},\Delta,\psi^{*},\psi]/\hbar\}\\
\nonumber&=\exp\Big\{\beta V\frac{|\Delta|^{2}}{V_{0}}+\text{Tr}[\log(-\mathbf{G}^{-1})]\Big\}\\
&=\exp\{-S^{\text{eff}}[\Delta^{*},\Delta]/\hbar\}.
\end{align}
where $\mathbf{G}^{-1}$ is
\begin{equation}
-\hbar\mathbf{G}^{-1}=\left(\begin{array}{cc}
\text{i}\hbar\omega_{n}-\hbar\omega_{+}(\mathbf{k})&0\\
0&\text{i}\hbar\omega_{n}-\hbar\omega_{-}(\mathbf{k})\end{array}\right).
\end{equation}

The interaction $V_{0}$ in the effective action can be eliminated
in favor of the two-body transition matrix using Eq.~(\ref{t2b bij
puntinteractie}). It is straightforward to calculate the
mean-field thermodynamic potential from the effective action,
which we find to be
\begin{align}
\nonumber&\omega_{L}(|\Delta|)=-\frac{1}{\beta V}\log Z=\\
\nonumber&-\frac{|\Delta|^{2}}{T^{\text{2B}}(0)}+\int\frac{\text{d}\mathbf{k}}{(2\pi)^{3}}\Big\{\varepsilon(\mathbf{k})-\mu-\hbar\omega(\mathbf{k})\\
&+\frac{|\Delta|^{2}}{2\varepsilon(\mathbf{k})}-\frac{1}{\beta}\sum_{\sigma=\pm}\log\left(1+e^{-\beta\hbar\omega_{\sigma}(\mathbf{k})}\right)\Big\},
\end{align}
where to obtain this thermodynamic potential the sum over the
Matsubara frequencies was performed using the following identity
\begin{equation}
\sum_{n}\log(\beta(-\text{i}\hbar\omega_{n}+\varepsilon))=\log\left(1+e^{-\beta\varepsilon}\right),
\end{equation}
which can be derived using contour integration.

\section{Feynman Diagrams}

To determine the critical temperature at which a phase transition
will take place we can, as mentioned in Section II, either
calculate coefficients from the thermodynamic potential by
deriving it with respect to $|\Delta|$ or we can use Feynman
diagrams. We present the latter scheme in some detail here.

From the action in Eq.~(\ref{actie voor diagrammen}) we can
determine the propagator for the fermionic fields and the vertices
for the interactions between the fermions and the bosons. Namely
the propagator in momentum space is
\begin{equation}
G_{0,\sigma}(\mathbf{k},\text{i}\omega_{n})=\frac{-\hbar}{-\text{i}\hbar\omega_{n}+(\varepsilon_{\sigma}(\mathbf{k})-\mu_{\sigma})}.
\label{green}
\end{equation}
This propagator is in the Feynman diagrams in
Figs.~\ref{ladderq}a and \ref{ladderq}b represented by a straight line,
where the plus (minus) indicates a light (heavy) particle. The
interaction vertex is proportional to
$\delta(\mathbf{k_{1}+k_{2}+k_{3}})\delta_{n,n',n''}$, where
$\mathbf{k}_{i}$ are the momenta and $n$ determine the frequencies
of the incoming and outgoing particles. This represents nothing
but conservation of momentum and energy. In the diagrams in
Figs.~\ref{ladderq}a and \ref{ladderq}b the vertices are the points where
three propagators meet.

In order to determine the Lifshitz point we need an expression for
the ladder diagram with nonzero external momenta.
\begin{align}
\nonumber&\frac{1}{\hbar\beta}\sum_{n}\int\frac{\text{d}\mathbf{k}}{(2\pi)^{3}}G_{0,+}(\mathbf{q}-\mathbf{k},-\text{i}\omega_{n})G_{0,-}(\mathbf{k},\text{i}\omega_{n})\\
\nonumber&=\frac{1}{\hbar\beta}\sum_{n}\int\frac{\text{d}\mathbf{k}}{(2\pi)^{3}}\frac{-\hbar}{\text{i}\omega_{n}\hbar+(\varepsilon_{+}(\mathbf{q-k})-\mu_{+})}\\
&\times\frac{-\hbar}{-\text{i}\omega_{n}\hbar+(\varepsilon_{-}(\mathbf{k})-\mu_{-})}.
\end{align}
We can split the fractions and then perform the summation over the
Matsubara frequencies. This results in
\begin{align}
\nonumber&\int\frac{\text{d}\mathbf{k}}{(2\pi)^{3}}\frac{N_{+}(\mathbf{q-k})+N_{-}(\mathbf{k})-1}{\varepsilon_{+}(\mathbf{q-k})+\varepsilon_{-}(\mathbf{k})-2\mu}.
\end{align}
If we want to obtain the full expression for the quadratic part,
we have to add the terms proportional to $|\Delta|^{2}$ from the
effective action. By doing so we obtain $\alpha(\mathbf{q})$ in
Eq.~(\ref{extern}). In the same way the diagrams can be calculated
needed for $\alpha$ in Eq.~(\ref{alpha}) and for $\beta$ in
Eq.~(\ref{beta}).

When we include fluctuation effects, we also take the screening of
the interaction by the particle-hole fluctuations into account. We
do so by replacing the bare interaction potential $V_{0}$ by a
screened interaction potential $V_{\text{sc}}$ containing the
infinite sum of so-called RPA bubble diagrams, see
Fig.~\ref{bubble}. The sum over all the bubble diagrams in
Fig.~\ref{bubble} is a geometric series. Using this, we find the result for the screened
interaction Eq.~(\ref{vsc}). Thus, to find the screened interaction, we
have to calculate the amplitude of the bubble diagram. The bubble
diagram consists of two fermionic propagators with momentum and
energy going in opposite directions through the diagram. The
amplitude is given by
\begin{align}
\nonumber\frac{1}{\hbar\beta}\sum_{n}\int\frac{\text{d}\mathbf{k}}{(2\pi)^{3}}&\frac{-\hbar}{-\text{i}\hbar\omega_{n}+\varepsilon_{+}(\mathbf{k})-\mu_{+}}\\
&\times\frac{-\hbar}{-\text{i}\hbar\omega_{n}+\varepsilon_{-}(\mathbf{k})-\mu_{-}}.
\end{align}
Writing the above expression as the sum of two fractions
and then performing the sum over the Matsubara frequencies results in
\begin{align}
\hbar\Pi(\mathbf{0},0)=\int\frac{\text{d}\mathbf{k}}{(2\pi)^{3}}\frac{N_{+}(\mathbf{k})-N_{-}(\mathbf{k})}{2h-(\varepsilon_{+}(\mathbf{k})-\varepsilon_{-}(\mathbf{k}))}.
\end{align}


\begin{thebibliography}{99}
\bibitem{Greiner} M.~Greiner, C.~A.~Regal, D.~S.~Jin, Nature {\bf 426}, 537 (2003).

\bibitem{Jochim} S.~Jochim {\it et al}., Science {\bf 302}, 2101 (2003).

\bibitem{Regal}C.~A.~Regal, M.~Greiner, D.~S.~Jin, Phys. Rev. Lett. {\bf 92}, 040403 (2004).

\bibitem{Zwierlein} M.~W.~Zwierlein {\it et al}., Phys. Rev. Lett. {\bf 92}, 120403 (2004).

\bibitem{Kinast} J.~Kinast {\it et al}., Phys. Rev. Lett. {\bf 92}, 150402 (2004).

\bibitem{Bartenstein} M.~Bartenstein {\it et al}., Phys. Rev. Lett. {\bf 92}, 203201 (2004).

\bibitem{Bourdel}T.~Bourdel {\it et al}., Phys. Rev. Lett. {\bf 93}, 050401 (2004).

\bibitem{Partridge}G.~B.~Partridge {\it et al}., Phys. Rev. Lett. {\bf 95}, 020404 (2005).

\bibitem{Ketterle} M.~W.~Zwierlein {\it et al}., Science {\bf 311}, 492 (2006).

\bibitem{Hulet} G.~B.~Partridge {\it et al}., Science {\bf 311}, 503 (2006).

\bibitem{Shin} Y.~Shin {\it et al}., Nature {\bf 451}, 689 (2008).

\bibitem{Gubbels} K.~B.~Gubbels, M.~W.~J.~Romans, H.~T.~C.~Stoof, Phys. Rev. Lett. {\bf 97}, 573    210402 (2006).

\bibitem{Diederix} J.~M.~Diederix, K.~B.~Gubbels, H.~T.~C.~Stoof, arXiv:0907.0127v2
(2009).

\bibitem{Bailin} D.~ Bailin and A.~Love, Phys. Rept. {\bf107}, 325 (1984).

\bibitem{Casalbuoni} R.~Casalbuoni and G.~Nardulli, Rev. Mod. Phys. {\bf 76}, 263 (2004).

\bibitem{prl} K.~B.~Gubbels, J.~E.~Baarsma, H.~T.~C.~Stoof, Phys. Rev. Lett. {\bf 103}, 195301 583    (2009).

\bibitem{Larkin} A.~I.~Larkin and Y.~N.~Ovchinnikov, Zh. Eksp. Teor. Fiz. {\bf 47}, 1136 (1964).

\bibitem{Fulde} P.~Fulde and R.~A.~Ferrell, Phys. Rev. {\bf 135}, A550 (1964).

\bibitem{Bulgac} A.~Bulgac and M.~McNeil Forbes, Phys. Rev. Lett. {\bf 101}, 215301 (2008).

\bibitem{Liao} Y.~Liao {\it et al}.,  arXiv:0912.0092 (2009).

\bibitem{Mora} C.~Mora and R.~Combescot, Phys. Rev. B {\bf 71}, 214504 (2005).

\bibitem{Yip} N.~Yoshida and S.-K.~Yip, Phys. Rev. A {\bf 75}, 063601 (2007).


\bibitem{Walraven} E.~Wille {\it et al}, Phys. Rev. Lett. {\bf 100}, 053201 (2008).

\bibitem{Dieckmann} A.-C.~Voigt {\it et al}, Phys. Rev. Lett. {\bf 102}, 020405 (2009).

\bibitem{Parish} M.~M.~Parish {\it et al}, Phys. Rev. Lett. {\bf 98} 160402
(2007).

\bibitem{Gorkov} L.~P.~Gor'kov and T.~K.~ Melik-Barkhudarov, Sov. Phys. JETP {\bf 13}, 1018
(1961).

\bibitem{Koos} K.~B.~Gubbels and H.~T.~C.~Stoof, Phys. Rev. Lett. {\bf 100}, 140407 (2008).

\bibitem{Burovski} E.~Burovski {\it et al}., Phys. Rev. Lett. {\bf 96}, 160402 (2006).

\bibitem{Duine} R.~A.~Duine and H.~T.~C.~Stoof, Phys. Rep. {\bf 396},115 (2004).

\bibitem{Romans} M.~W.~J.~Romans, H.~T.~C.~Stoof, Phys. Rev. Lett. {\bf 95}, 260407 (2005).

\bibitem{Negele} J.~W.~Negele, H.~Orland, {\it Quantum Many-Particle Systems} (Westview Press, Boulder, 1998).

\bibitem{uqf} H.~T.~C.~Stoof, K.~B.~Gubbels, D.~B.~M.~Dickerscheid, {\it Ultracold Quantum Fields} (Springer, Dordrecht, 2009).

\bibitem{Chaikin} P.~M.~Chaikin and T.~C.~Lubensky, {\it Principles of Condensed Matter Physics} (Cambridge University Press, Cambridge, 1995).

\bibitem{Iskin} M.~Iskin and C.~A.~R.~S\'a de Melo, Phys. Rev. A {\bf 77}, 013625 (2008).

\bibitem{Fetter} A.~L.~Fetter, J.~D.~Walecka, {\it Quantum Theory of Many-Particle Systems} (McGraw-Hill, New York, 1971).

\bibitem{Sarma} G.~Sarma, J. Phys. Chem. Solids {\bf 24}, 1029 (1963).

\bibitem{Kleinert} H.~Kleinert, Forts. Phys. {\bf 26}, 565 (1978).

\bibitem{Paananen} T.~Paananen, P.~T\"orm\"a, J.-P.~Martikainen, Phys. Rev. A {\bf 75}, 023622 (2007).

\bibitem{Marchetti} M.~M.~Parish {\it et al}, Nature Physics {\bf 3}, 124 (2007).

\bibitem{Buzdin} A.~I.~Buzdin and H.~Kachkachi, Phys. Lett. A {\bf 225}, 341 (1997)

\bibitem{Melo} C.~A.~S\'{a} de Melo {\it et al}., Phys. Rev. Lett. {\bf 71}, 3202 (1993).

\bibitem{Combescot} R.~Combescot {\it et al}., Phys. Rev. Lett. {\bf 98},
180402 (2007).

\bibitem{Lobo} C.~Lobo {\it et al}., Phys. Rev. Lett. {\bf 97}, 200403 (2006).

\bibitem{Carlson} A.~Gezerlis {\it et al}., arXiv:0901.3105
(2009).


\end{thebibliography}
\end{document}